\documentclass[oldversion]{aa} 
\usepackage{natbib}
\usepackage{graphicx}
\begin{document}
\title{$uvby-H_{\beta}$ CCD photometry and membership segregation of the 
open cluster NGC~2682 (M~67).
\thanks{Tables~3, 4 and 7 are only available in electronic form from 
CDS via anonymous ftp to cdsarc.u-strasbg.fr (130.79.128.5) or
via http://cdsweb.u-strasbg.fr/Abstract.html}
    }

   \author{L. Ba\-la\-guer-\-N\'u\-\~nez\inst{1,2,3}, 
   D. Galad\'{\i}-Enr\'{\i}quez\inst{4},
   C. Jor\-di\inst{1,5} 
          }

\offprints{Balaguer-N\'u\~nez, L., \email{Lola.Balaguer@am.ub.es}}

\institute{Departament d'Astronomia i Meteorologia, Universitat de
    Barcelona, Avda. Diagonal 647, E-08028 Barcelona, Spain 
 \and
    Shanghai Astronomical Observatory, CAS Shanghai 200030,
	P.R. China
 \and
    Institute of Astronomy, 
    Madingley Road, CB3 OHA Cambridge, UK
 \and         
    Instituto de Astrof\'{\i}sica de Andaluc\'{\i}a (CSIC). 
    Camino Bajo de Hu\'etor 50, E-18008 Granada, Spain
\and
    Institut d'Estudis Espacials de Catalunya - IEEC, Edif. Nexus
    Gran Capit\`a 2-4, E-08034 Barcelona, Spain
     }
\date{Received 22 December 2006 / Accepted 13 April 2007}

\authorrunning{Balaguer-N\'u\~nez et al}
\titlerunning{$uvby-H_{\beta}$ Photometry and Membership of M~67}

\abstract{
Following deep astrometric and photometric study of the cluster 
NGC~2682 (M~67), we are able to accurately determine its fundamental 
parameters.  Being an old and rich cluster, M~67 is a relevant 
object for the analysis of the Galactic disk evolution.
M~67 is well studied but the lack of a wide and deep 
Str\"omgren photometric study makes our results worthwhile. 
The brightest stars of the open cluster M~67 were used as 
$uvby-H_{\beta}$ standard stars in our studies of NGC~1817
and NGC~2548, and the extension of the field covered, as well as the
amount of observations, allowed to obtain the best set of 
Str\"omgren data ever published for this cluster.
We discuss the results of our CCD $uvby-H_{\beta}$ intermediate-band 
photometry, covering an area of about 50$\arcmin \times$50$\arcmin$ 
down to $V \sim$ 19. Moreover, a complete membership segregation 
based on astrometric and photometric criteria is obtained. 
The photometric analysis of a selected sample of stars yields a
reddening value of $E(b-y)$~= 0.03$\pm$0.03, a distance 
modulus of $V_0-M_V$ = 9.7$\pm$0.2 and [Fe/H] = 0.01$\pm$0.14.
Through isochrone fitting we found an age of $\log t$ = 9.6$\pm$0.1 
(4.2$\pm$0.2~Gyr).  A clump of approximately 60 stars around 
$V$ = 16, $(b-y)$ = 0.4 could be interpreted as a population 
of pre-cataclysmic variable stars (if members), or as a stream of 
field G-type stars placed at twice 
the distance of the cluster (if non-members).
}

\maketitle


\section{Introduction}
NGC~2682 (C0847+120), also known as M~67, in Cancer 
[$\alpha_{2000}$=8$^{\mathrm h}51^{\mathrm m}\llap{.}3$,
$\delta_{2000}=+11{\degr}50\arcmin$; $l~= 215^\circ\llap{.}66$, 
$b~= +31^\circ\llap{.}91$] is probably the most 
thoroughly studied old open cluster
in the Galaxy, thanks to its small distance from us 
(estimated to be $\sim$ 900~pc).  Typically quoted
values for the age of the cluster ($\sim$ 4~Gyr) place it among the 
oldest open clusters.  In our astrometric and photometric 
study of several open clusters \citep{Bal04b,Bal05},
the photometric standard stars were taken from this cluster \citep{Nissen}, 
thus high-quality Str\"omgren wide-field CCD photometry of it also 
resulted from our observations.
The existence of a proper motions study \citep{zhati} of similar quality 
and characteristics of those used in our study of other clusters
(based on the same plate material and methods, by the Shanghai 
Astronomical Observatory), allowed us to apply 
the same techniques and analysis tools
to this cluster. This paper on M~67 closes the series at this stage of our open 
cluster programme. 

Photometric studies of M~67 have been performed by 
several authors. \cite{Mont}
have presented a deep ($V \sim$ 20) $UBVI$ CCD photometry of 1468 stars
within 15$\arcmin$ of the centre of the cluster, and \cite{Fan} have studied
spectrophotometry of similar depth in nine BATC intermediate-band filters
for stars in a 1$\fdg$92 $\times$ 1$\fdg$92 area centred on the cluster.
In addition, variability studies have been performed in this cluster, most 
notably by \cite{Gilli}, who conducted a very sensitive, highly 
temporally sampled study of stars in the central few arcmin of the cluster. 
\cite{Stassun} using differential CCD photometry to search for
variability, obtained sensitive photometry of 990 stars in a roughly
square region one-third of a degree on a side centred 5$\arcmin$ 
north of the cluster centre.  \cite{Sand} also as a by-product of 
variability studies, conducted a 
high precision $VI$ colour-magnitude diagram analysis.

\cite{Nissen} obtained accurate $uvby-H_{\beta}$ photoelectric 
photometry of a sample of 79
stars within a radius of 10$\arcmin$ from the centre, from the subgiant branch 
down to the unevolved main sequence. For this reason 
stars in M~67 were used by us as photometric standard stars for the
observations of NGC~1817 and NGC~2548 
\citep{Bal04b,Bal05}.
The amount of observations taken in our programme allowed us to obtain  
high quality photometry of this cluster. 
In spite of being a well studied cluster, 
our Str\"omgren photometry deserves an analysis due both to its 
deepness and to the extension of the covered area.

There are also a number of proper motion and radial velocity studies. 
\cite{Sanders77} calculated probabilities of membership based on
relative proper motions of 1866 stars in the cluster field with a 
limiting photographic magnitude of $\sim$17. From 10 plate pairs 
of different origin and with a
maximum epoch difference of 68 years, he found 649 probable members. 	
\cite{Girard} gave relative proper motions for 663 stars in an area
of 42$\arcmin \times$34$\arcmin$ from 44
plates with a maximum separation of 66 years. Although the plates used
were of different depths, the deepest plates provided a limiting visual 
magnitude down to 16. 
In 1993, Zhao et al. derived relative proper motions for 1046 stars 
within a 1$\fdg$5 $\times$
1$\fdg$5 area in the region from PDS measurements  
of 9 plates with a maximum epoch difference of 80 years and a magnitude
limit $V \sim$ 15.5. 

On the other hand, the radial velocity studies by Mathieu et al.\ 
(\citeyear{Mathieu}, \citeyear{Mathieu90}) gave precise radial-velocity 
measurements for 170 stars, including all main-sequence stars brighter 
than $V$~= 12.8.


In this paper we discuss the results of our CCD photometric study, 
covering an area of 
about 50$\arcmin$$\times$50$\arcmin$ (Fig.~\ref{map67}) 
down to $V$$\sim$ 19.  
Section~2 contains the details of the CCD observations and their 
reduction and transformation to the standard system. In Sect.~3 we discuss
a new membership segregation based on the combination of
parametric and non-parametric methods applied to the above-mentioned
proper motions from Zhao et al.\ (\citeyear{zhati}). 
This study was selected among all other existent because it covers
the biggest area and largest epoch difference.
Also, this material is fully homogeneous with that used by us for the other 
clusters in this series, as from the point of view of telescope and plates,
as from the point of view of proper motion treatment and reduction. 
This section includes a discussion on the limits on proper motions detection
of binarity. 

In Sect.~4 we discuss the colour-magnitude diagram and 
identify the sample of probable cluster members
using astrometric as well as photometric criteria. 
Section~5 contains the derivation of the fundamental cluster
parameters of reddening, distance, metallicity and age. 
Section~6 studies the multiple star systems, 
blue stragglers and a new feature on the colour-magnitude diagram.
Section~7 summarizes our
conclusions. 


\section{The Data}

\subsection{Observations}

The Str\"omgren CCD photometry of the area was obtained at Calar 
Alto Observatory (Almer\'{\i}a, Spain) in January 2000
using the 1.23 m telescope of Centro Astron\'omico Hispano-Alem\'an (CAHA).  
Further data were obtained at Observatorio del Roque de los Muchachos 
(ORM, La Palma, Canary Islands, Spain) in February 2000 using the 2.5 m 
Isaac Newton Telescope (INT) of ING (equipped with the Wide-Field Camera, WFC), 
and in December 1998 and February 2000 using the 1 m Jakobus Kapteyn 
Telescope (JKT) of ING, with the $H_{\beta}$ filters.
A log of the 
observations, the total number of frames,
exposure times and seeing conditions is given in Table~\ref{log67}.

\begin{table*}
\leavevmode
\caption {Log of the observations used in this study.}
{\footnotesize
\begin{center}
\begin {tabular} {ccccccccc}
\hline
Telescope  & Date & Seeing($\arcsec$) & N. of frames  & \multicolumn{5}{c}{Exp. Times (s)} \\
&  &  &  &  $u$ & $v$ & $b$ & $y$ & $H_\beta$  \\
\hline

1.23 m CAHA  & 2000/01/05-10 & 1.1 & 20 &  &  & &  & 90 \\
1 m JKT &  2000/02/02-06 & 1.1 & 42 & - & - & - & - & 200 \\
2.5 m WFC-INT & 2000/02/02-03 & 1.3 & 44 & 80 & 20 & 20 & 10 & - \\

\hline
\label{log67}
\end {tabular}
\end {center}
}
\end {table*}

\begin{figure}
\resizebox{\hsize}{!}{\includegraphics{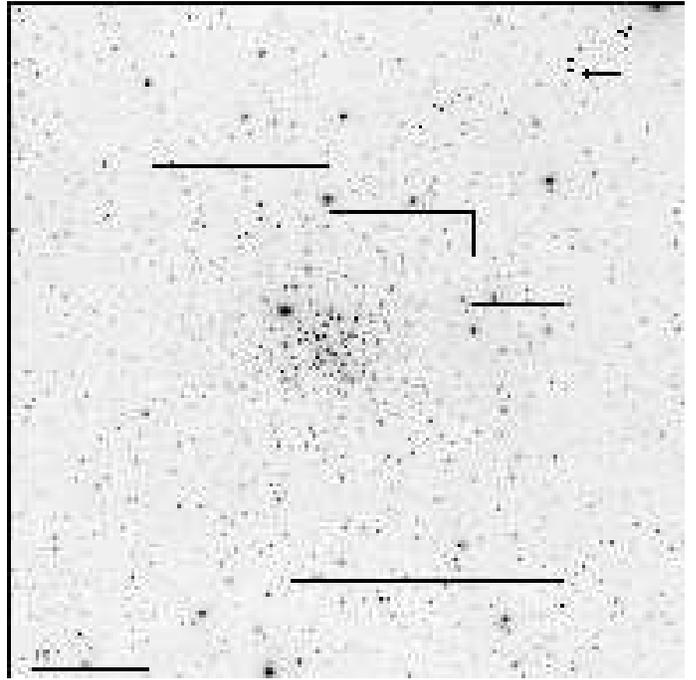}}
\caption{Finding chart of the area under study. The covered area 
is marked in black on an 
image of a plate (POSSI.E -optical R-.DSS1.486-LOW) plotted with 
Aladin \citep{Aladin}.
The $H_{\beta}$ spatial coverage is limited to the 
central region. See text for details.}
\label{map67}
\end{figure}

We obtained photometry for a total of 1843 stars in an area of 
50$\arcmin$$\times$50$\arcmin$ in M~67 region, down to a 
limiting magnitude $V$$\sim$ 19.
The area covered is shown in the finding chart 
of the cluster (Fig.~\ref{map67}).
Due to the lack of $H_{\beta}$ filter at the WFC-INT, it was only possible to
measure it at the JKT and CAHA telescopes, thus limiting the spatial coverage
with this filter.
Only 288 stars (in the central region) have $H_{\beta}$ values.

\subsection{Reduction and Transformation}

The reduction of the photometry is explained in length at 
Balaguer-N\'u\~nez et al. (\citeyear{Bal04b}). 
Having the standard stars in the same images,
the process is, in this case, equivalent to performe differential 
photometry. 

Our general procedure has been to routinely obtain twilight 
sky flats for all the filters 
and a sizeable sample of bias frames (around 10) before and/or after every run. 
Flat fields are typically fewer in number, from five to ten per filter.  
Two or three dark frames of 2000~s were also taken. 
IRAF\footnote{IRAF is distributed by the National Optical Astronomy 
Observatories,
which are operated by the Association of Universities for Research
in Astronomy, Inc., under cooperative agreement with the National
Science Foundation.} routines were used for the reduction process.
The bias level was evaluated individually for each frame by averaging the counts
of the most stable pixels in the overscan areas. The 2-D structure of the bias
current was evaluated from the average of a number of dark frames with 
zero exposure time.  Dark current was found to be negligible in all the cases.
Flatfielding was performed using 
sigma clipped, median stacked, dithered twilight flats.
Ten short exposures in every 
filter were taken every night with a magnitude limit of $V$$\sim$ 19. 
At least one exposure per filter was pointed to make the centre of the 
cluster fall on each of the WFC chips.

Our fields are not crowded. Thus, the synthetic aperture
technique provides the most efficient measurements of relative fluxes within 
the frames and from frame to frame. We used the appropiate IRAF packages, 
and DAOPHOT and DAOGROW algorithms \citep{Stet87, Stet90}. 
We analyzed the magnitude growth curves and determined the aperture correction 
with the IRAF routine MKAPFILE. 

For WFC images from the INT, we employ  the pipeline specifically 
developed by the Cambridge Astronomical Survey Unit. 
The process bias subtracts, gain corrects and flatfields the images.
Catalogues
are generated using algorithms described in Irwin (\citeyear{Irw}).
The pipeline gives accurate positions 
in right ascension and declination linked to the USNO2 Catalogue 
\citep{USNO2}, and instrumental magnitudes with 
their corresponding
errors. A complete description can be found in Irwin \& Lewis (\citeyear{Irwin})
and in {\tt http://www.ast.cam.ac.uk/\~{}wfcsur/index.php}.

\begin{table*}
\caption {Number of stars observed ($N$) and mean internal errors 
   ($\sigma$) as a function of apparent visual magnitude.}
\begin{center}
\begin {tabular} {ccccccccccc}
\hline
\multicolumn{1}{c}{$V$ range} & 
\multicolumn{2}{c}{$V$} & 
\multicolumn{2}{c}{$(b-y)$} & 
\multicolumn{2}{c}{$m_1$} & 
\multicolumn{2}{c}{$c_1$} & 
\multicolumn{2}{c}{$H_{\beta}$} \\
\hline
& $N$&$\sigma$ & $N$&$\sigma$ & $N$&$\sigma$ & $N$&$\sigma$ &$N$&$\sigma$ \\
\hline
8- 9 &    3 & 0.029 &    3 & 0.028 &    3 & 0.029 &    2 & 0.004 &      &        \\
9-10 &    8 & 0.008 &    7 & 0.007 &    6 & 0.008 &    7 & 0.007 &    2 & 0.001  \\
10-11 &   22 & 0.003 &   22 & 0.002 &   22 & 0.003 &   22 & 0.005 &   10 & 0.009  \\
11-12 &   28 & 0.004 &   28 & 0.022 &   26 & 0.003 &   26 & 0.005 &   14 & 0.016  \\
12-13 &  114 & 0.004 &  114 & 0.003 &  114 & 0.004 &  114 & 0.005 &   56 & 0.009  \\
13-14 &  199 & 0.003 &  199 & 0.003 &  198 & 0.004 &  197 & 0.006 &   75 & 0.015  \\
14-15 &  240 & 0.020 &  239 & 0.005 &  238 & 0.006 &  237 & 0.009 &   63 & 0.016  \\
15-16 &  277 & 0.007 &  277 & 0.010 &  275 & 0.011 &  269 & 0.018 &   57 & 0.019  \\
16-17 &  299 & 0.011 &  299 & 0.015 &  272 & 0.021 &  233 & 0.031 &   10 & 0.016  \\
17-18 &  307 & 0.021 &  306 & 0.029 &  245 & 0.041 &  109 & 0.043 &    1 & 0.033  \\
18-19 &  259 & 0.040 &  259 & 0.058 &  130 & 0.073 &   46 & 0.059 &      &        \\
19-20 &   75 & 0.093 &   75 & 0.125 &   18 & 0.141 &   12 & 0.106 &      &        \\
\hline
Total & 1831 &       & 1828 &       & 1547 &       & 1274 &       &  288 &        \\
\hline
\label{error67}
\end {tabular}
\end{center}
\end {table*}

\begin{figure}
\begin{center}
\resizebox{\hsize}{!}{\includegraphics{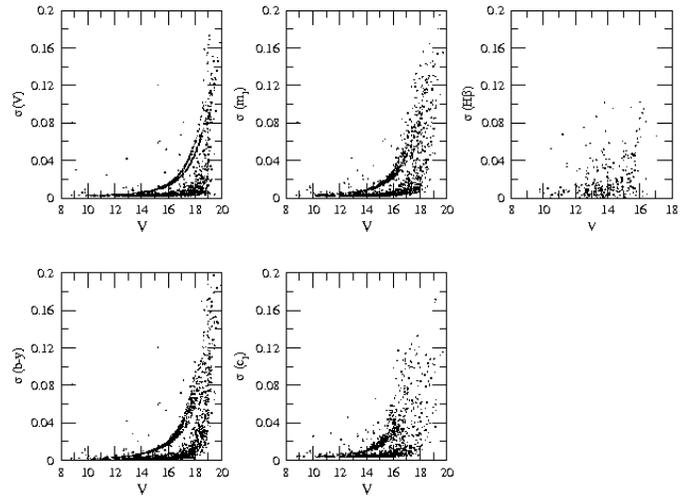}}
\caption{Mean internal errors of magnitude and colours (including the 
effects both from the instrumental measurements and the transformation 
uncertainties) 
as a function of the apparent visual magnitude, $V$, 
for all observed stars in the cluster region. The
structure in the magnitude dependence is mainly owed to different numbers of
measurements for individual stars since the centre of the cluster was shifted
to be observed by each of the four CCD chips. 
}
\label{errby67}
\end{center}
\end{figure}

The coefficients of the transformation equations were computed 
by a least squares method using the instrumental magnitudes of the
standard stars and the standards magnitudes and colours in the 
$uvby-H_{\beta}$ system. Up to 68 standard stars in the field of the cluster 
(Nissen et al.\ \citeyear{Nissen}) were used depending on 
the size of the frame. 
Those standard stars with residuals greater than 
2$\sigma$ were rejected. 
The reduction was performed for each night independently and in two steps.
The first step is to determine the extinction coefficients for
each passband from the standard stars. With the extinction coefficients 
fixed, the transformation coefficients to the standard system 
were fitted.
The final errors as a function of apparent visual magnitude
are given in Table~\ref{error67}
and plotted in Fig.~\ref{errby67}.

Table~3 
lists the $u,v,b,y,H_{\beta}$ data for all 1843 stars observed in a
region of 50$\arcmin$$\times$50$\arcmin$ around the
open cluster M~67 (Fig.~\ref{map67}). Star positions 
are given as frame ($x,y$)
and equatorial ($\alpha_{\mathrm J2000}$,$\delta_{\mathrm J2000}$) coordinates. 
An identification number was assigned to each star following 
the order of increasing right ascension.
Column 1 is the ordinal star number; cols. 2 and 3
are $\alpha_{\mathrm J2000}$ and $\delta_ {\mathrm J2000}$;
cols. 4 and 5 are the respective $x$, $y$ coordinates in arcmin
(arbitrary origin, close to cluster centre);
cols. 6 and 7 are the $(b-y)$ and its error,
cols. 8 and 9 the $V$ magnitude and its error, 
cols. 10 and 11 the $m_1$ and its error, 
cols. 12 and 13 the $c_1$ and its error, 
and cols. 14 and 15 the $H_{\beta}$ and its error.
In col. 16, stars considered candidate members (see Sect.~4.1.) are labelled 
'M', while those classified as non-members show the label 'NM'.

\addtocounter{table}{1}

The cross-identification of stars in common 
with the astrometry \citep{zhati}, BDA
({\tt htpp://obswww.unige.ch/WEBDA}), 
\cite{Sanders77},
Hipparcos \citep{esa}, Tycho-2 \citep{tyc2} 
and USNO-2 \citep{USNO2} catalogues
is provided in Table~4.


\addtocounter{table}{1}

\subsection{Comparison with Previous Photometry}

After \cite{Nissen}, few studies of this cluster have been done in Str\"omgren 
photometry. 
\cite{Anthony} used a CCD to cover a larger area 
with up 
to 44 stars in common with our sample. Comparing \citeauthor{Anthony}'s
data with ours, the mean differences in the sense ours minus others, 
we get $-$0.01($\sigma$~= 0.04) in $V$,
0.00(0.02) in $b-y$, 0.00(0.04) in $m_1$ and 0.00(0.04) in $c_1$.  
The data published by \cite{Joner}, with 14 stars in common with our catalogue,
up to a limit of $V \sim$ 13, give differences of 
0.01(0.03) in $V$, 0.01(0.02) in $b-y$, $-$0.02(0.03) in $m_1$ and
$-$0.01(0.04) in $c_1$. 
Before \cite{Nissen}, the study of \cite{StromStrom} with 19 common stars 
up to a limit of $V \sim$ 13.5 gives:
$-$0.02(0.03) in $V$, $-$0.03(0.02) in $b-y$, 0.03(0.05) in $m_1$,
$-$0.02(0.06) in $c_1$ and $-$0.02(0.07) in $H_{\beta}$, but a large scatter
and a probable colour term in their results was already noted by \cite{Nissen}.

The $V$ magnitude derived from the $y$ filter
can be compared to the published broadband data.
Several studies of M~67 give us
mean differences in $V$, in the sense ours minus others (see Table~\ref{CompV67}).
Looking at the differences, one concludes that there are no apparent systematic trends in 
our photometry.

\begin{table}
\leavevmode
\caption {Comparison of the $V$ magnitude to the published broadband data. 
}
{\tiny
\begin{center}
\begin {tabular} {lcr@{.}lcr}
\hline
Author & &\multicolumn{2}{c}{$\Delta V_{\rm our-oth}$} & $\sigma$ & $N$ \\
\hline
\cite{Eggen}  & photoel.& $-$0&02 & 0.04 & 173 \\
\cite{Sanders}& photoel.&    0&01 & 0.05 & 225 \\
\cite{Gilli}  & CCD     & $-$0&02 & 0.03 & 136 \\   
\cite{Mont}   & CCD     & $-$0&01 & 0.05 & 967 \\
\cite{Kim}    & CCD     & $-$0&01 & 0.04 & 89 \\ 
\cite{Hend}   & CCD     &    0&01 & 0.06 & 306 \\    
\cite{Sand}   & CCD     &    0&01 & 0.03 & 154 \\ 
\hline
\label{CompV67}
\end {tabular}
\end {center}
}
\end {table}


\section{Astrometric Analysis}

We have studied the membership segregation using parametric and 
non-parametric approaches in the same way as in \cite{Bal04a,Bal05}. 
Thanks to the proper 
motions obtained by \cite{zhati} from high-quality plates 
taken with the double astrograph at the Z\v o-S\`e station of the 
Shanghai Observatory, as in the case of NGC~1817 and NGC~2548, 
we have been able to use the same procedure in a very homogeneous way. 
The focal length of the Gautier 40~cm double astrograph at the 
Z\v o-S\`e station is 6.9~m (hence a plate scale 
of 30\arcsec~mm$^{-1}$). All the plates were measured using the Photometric
Data Systems model 1010 automatic measuring machine at the Purple Mountain
Observatory in Nanjing. 
\citeauthor{zhati} give relative proper motions of 
1046 stars within a 1$\fdg$5 $\times$ 1$\fdg$5 area,
from measurements of 9 plates. 
The plates have a maximum epoch difference of 80 years. 

\cite{zhati} applied
the plate-pair technique adopted many times 
at Shanghai Observatory to derive the relative proper motions.
All linear 
and quadratic coordinate-dependent terms and the coma term were included 
in the plate-pair solution. 
The quoted mean errors of the relative proper motions vary
from 0.4~mas~yr$^{-1}$ for bright stars in the inner part of the cluster 
field to
some 1.5~mas~yr$^{-1}$ for faint stars in the outer part of the cluster.
The comparison with the proper motions of \cite{Sanders77} and \cite{Girard}
shows a satisfactory agreement.
The magnitude limit of \citeauthor{zhati}'s study is $V \sim$ 15.5.

There are 50 stars in common with the Tycho-2 Catalogue \citep{tyc2}. 
From 
their absolute proper motions we can calculate the transformation of 
\citeauthor{zhati}'s relative proper motions ($\mu_{x},\mu_{y}$)
to the absolute reference frame:

{\centering
$(\mu_{\alpha}\cos\delta)_{\mathrm{TYC2}}$ = 
$-6.510 \ ({\pm} 0.259) + 1.013 \ ({\pm} 0.034) \cdot \mu_{x} - 0.017 \ ({\pm} 1.644) \cdot \mu_{y}$;\\
$r$~= $0.956$; $N$~= 44\\
$(\mu_{\delta})_{\mathrm{TYC2}}$~= 
$-8.051 \ ({\pm} 0.404) + 0.025 \ ({\pm} 0.056) \cdot \mu_{x}
+ 0.957 \ ({\pm} 0.056) \cdot \mu_{y}$;\\
$r$~= $0.951$; $N$~= 50 \\
}

\noindent  where $r$ is the correlation coefficient and proper motions are
expressed in mas~yr$^{-1}$. The transformations show a rather
good alignment of $x$- and $y$-axis with right ascension and
declination, respectively.

Membership determination in Zhao et al. (\citeyear{zhati}) was calculated
with an 8-pa\-ra\-me\-tric Gaussian model, and a list of stars with probability 
higher than 0.8 and a distance to the centre less than 45$\arcmin$ 
gave 282 cluster members.  
For the sake of coherence with our analysis of NGC~1817 and NGC~2548
\citep{Bal04a,Bal05}, 
we apply a 9-parametric Gaussian model and a non-parametric method 
to the proper motions 
obtained by \citeauthor{zhati} 
In contrast to 
their approach to the segregation of cluster members, 
we do not use any spatial information. 
Selecting members on kinematic and
photometric criteria leaves the spatial information untouched, ready for
a clean interpretation of radial and spatial trends.

\citeauthor{zhati} give $x,y$ coordinates, 
relative proper motions with their errors, number 
of plates used for proper motion determination,   
and cross-identifications with \cite{Sanders77}. 
The authors state that their $x,y$ coordinates are taken from one of the plates,
but a detailed analysis of the data shows that this is the case for the
majority of the stars, but there is a small subset (61 stars) whose $x,y$
coordinates seem to have been taken from at least two other plates,
and this introduces shifts in their quoted positions. Fortunately, 
\citeauthor{zhati}'s cross-identifications with \cite{Sanders77} were
correct, what allowed us a complete cross-identification of our
photometric catalogue with the proper motion data not losing any star.
This inhomogeneity in \cite{zhati} $x,y$ data does not affect our results, since we do
not use these coordinates in our study. All tests performed show 
that Zhao et al.'s 
proper motions are not affected. The mistake
in the $x,y$ coordinates of 61 stars seems to have been 
introduced in the
final preparation of the tables by \cite{zhati}.

\subsection{Membership probability: general frame}
\label{frame}

Let's consider any observational plane
($a, b$), where $a$ and $b$ stand for any physical magnitudes.
The distribution of any stellar sample on this plane can be described
by its frequency function, $\Psi(a,b)$, measured in units of number
of stars per unit area. The integral of the frequency function all over
the plane, i.e., its volume, gives the number of members of the
sample, $N$. 

If the frequency fuction is normalised to unit volume, 
then we get the probability density function (PDF), $\psi(a,b)=\Psi(a,b)/N$,
measured in units of fractions of the sample per unit area. 

If two populations are superposed on the same region of the
observational plane, but following different frequency functions
$\Psi_{1}$ and $\Psi_{2}$, then, for any point ($a,b$) on that plane,
it is possible to compute the probability $P$ for a star placed at that point
to belong to population $i\;\; (i=1,2)$: 
\begin{equation}
\label{probabilitat}
P = \frac{\Psi_{i}(a,b)}{\Psi_{1}(a,b)+\Psi_{2}(a,b)}. 
\end{equation}

For our purposes, ($a, b$) are the proper motions  
and the observational plane is then called
the vector-point diagram (VPD). 

\begin{table*}
\leavevmode
\caption{Distribution parameters and their uncertainties
from a 9-parametric Gaussian model applied to M~67 cluster and the field.
$n_{\rm c}$ measures the volume of the cluster frequency function 
($n_{\rm f} = 1- n_{\rm c}$), 
$\mu_x$ and $\mu_y$ are the mean relative proper motions, $\sigma_{\rm c}$,
and $\sigma_{\mu_x}$ and $\sigma_{\mu_y}$ are the dispersions of the
Gaussians fitted to the cluster and field distributions,
and $\rho$ is the correlation for the field function (it fixes the 
orientation of the field elliptical function).
The units of $\mu$ and $\sigma$ are mas~yr$^{-1}$.}
{\small      
\begin{center}
\begin {tabular} {lc c c c c c c c c }
\hline
& $n_{\rm c}$
& $ \mu_{x}$ &
$\mu_{y}$ & 
$\sigma_{\rm c}$  &
$\sigma_{\mu_{x}}$ & $\sigma_{\mu_{y}}$ 
& $\rho$\\
\hline
M~67& 0.364 & $-$0.59 & 0.49 & 0.89 &  &  &  \\
& $\pm0.016$ & $\pm0.07$ & $\pm0.06$ & $\pm0.05$  &  &  &  \\

field  &
& 0.02 & 3.31  &  &9.00 & 8.63 & $-0.21$\\
&   & $\pm0.10$ & $\pm0.48$ & &$\pm0.03$ & $\pm0.24$ & $\pm$0.02 \\
\hline
\label{para67}
\end {tabular}
\end {center}
}
\end {table*}

\subsection{The classical approach}
\label{parametric}

The parametric method assumes the existence of two populations 
in the VPD: cluster and field. The corresponding frequency functions
are modelled as parametric Gaussian functions: a circular Gaussian model
is adopted for the cluster distribution, while a bivariate (elliptical)
Gaussian is admitted to describe the field. We apply
a 9-parametric Gaussian model following \citeauthor{Bal04a} 
(\citeyear{Bal04a}, \citeyear{Bal05})
and the results are shown in Table~\ref{para67}.
Our implementation takes into account the individual errors of
the proper motion measurements \citep{zhaohe}.  If we transform the obtained 
mean relative proper motion of the cluster to the absolute 
system by the equations above, we get 
($\mu_{\alpha}\cos\delta, \mu_{\delta}$)~= ($-$7.1$\pm$0.8,$-$7.6$\pm$0.4) mas~yr$^{-1}$.

%

\begin{figure}
\begin{center}
\resizebox{6.0cm}{!}{\includegraphics{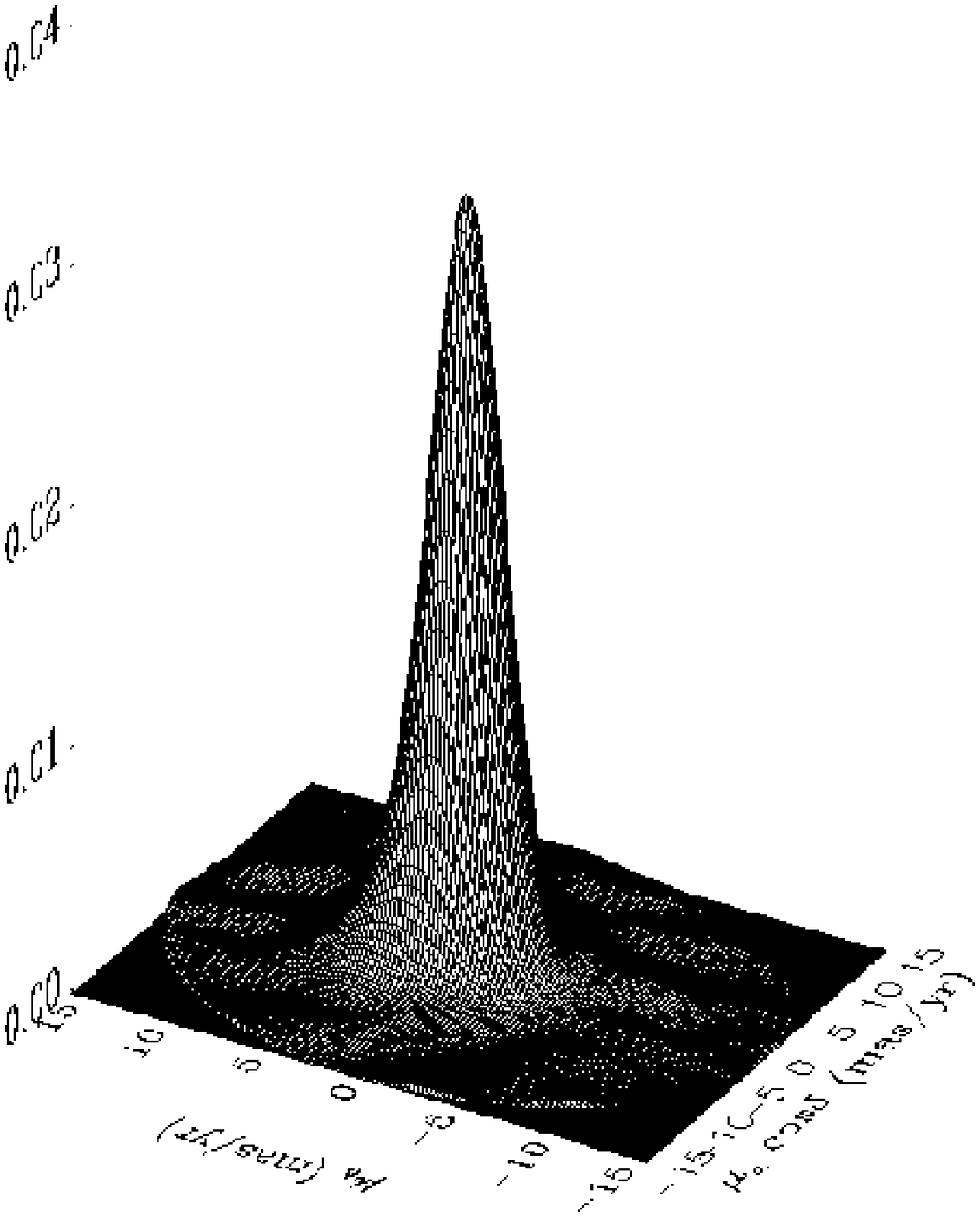}}\\
\resizebox{6.0cm}{!}{\includegraphics{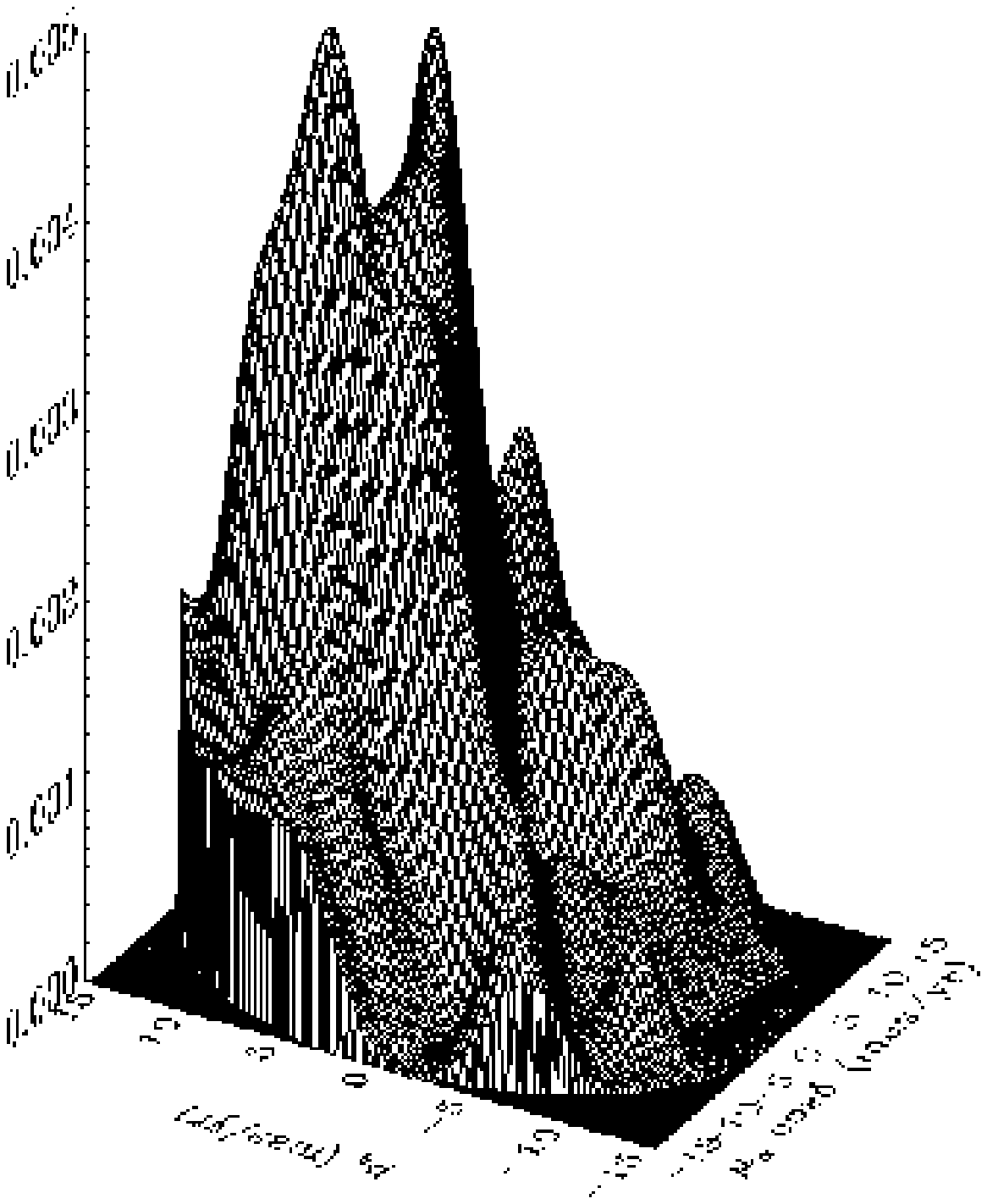}}\\
\resizebox{6.0cm}{!}{\includegraphics{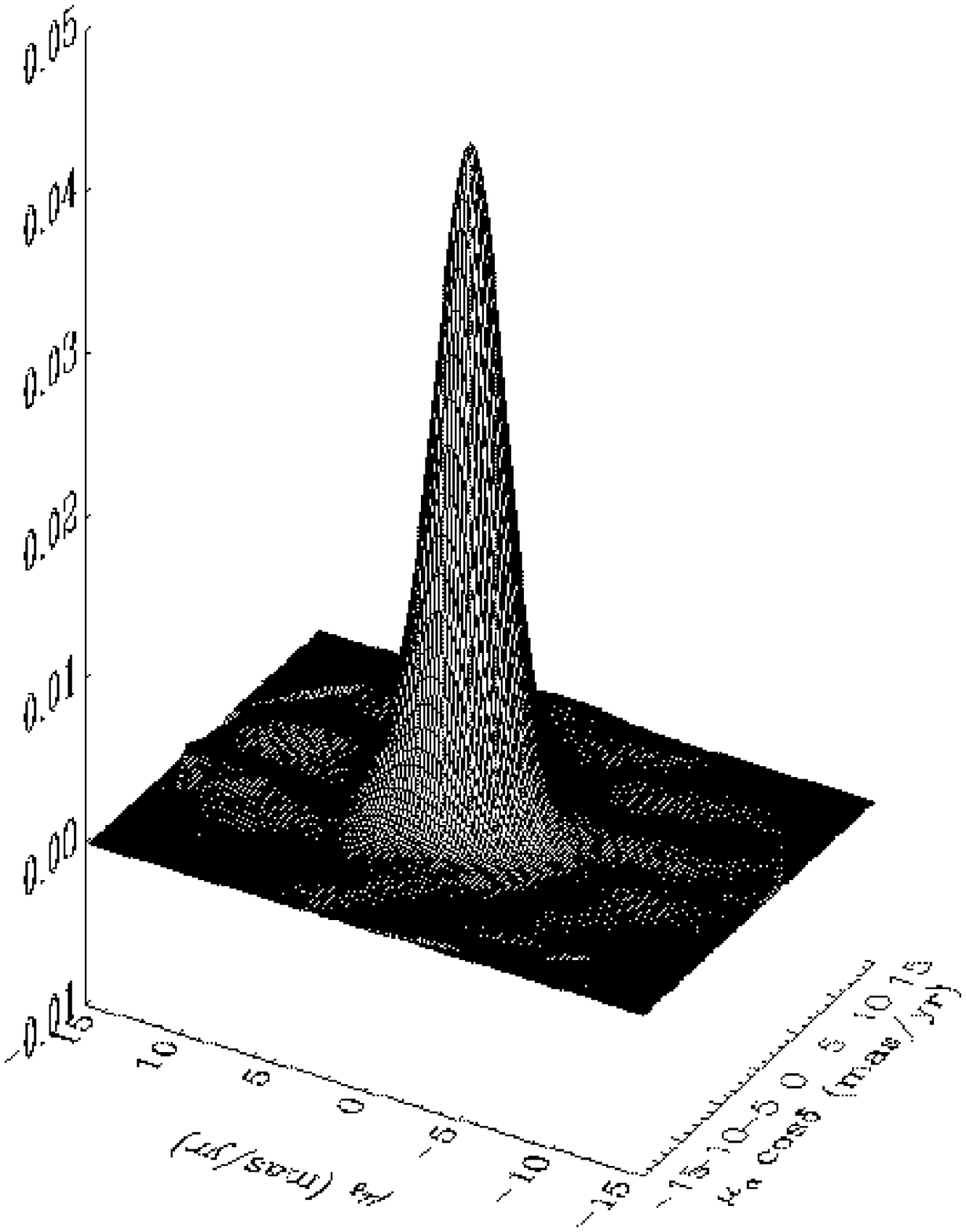}}
\caption{Empirical PDFs in the vector-point diagram.
Top: $\psi_{\rm c+f}$ mixed sample
from the inner circle of 35' centred on the point of maximum spatial
density of stars. Centre: $\psi_{\rm f}$ field population from 
outside this circle. Bottom: $\psi_{\rm c}$ cluster population of M~67.
It is noticeable that the cluster lies on a poorly populated background field.}
\label{nprob67}
\end{center}
\end{figure}

\subsection{The non-parametric approach}
\label{nonparametric}

The cluster/field segregation from astrometry has also been analysed
with a non-parametric approach, as explained in length in \cite{Bal04a, Bal05}.
We perform an empirical determination of the frequency functions
from the vector-point diagram (VPD),
without relying on any previous assumption about their profiles.
In the area occupied by the cluster, 
the frequency function $\Psi_{\rm c+f}$ is
made up of two contributions: cluster, $\Psi_{\rm c}$, and field, $\Psi_{\rm f}$.
To disentangle the two populations, we studied the VPD for the plate area
outside a circle centred on the cluster.
To this end, the centre of the cluster was chosen as the point
of highest spatial density.
The only two assumptions that we need to apply are (1) that it is possible to
determine the frequency function of the field stars from some spatial
area free of cluster members (outside the circle in our case); and, (2) that
this frequency function is representative of the field frequency function
in the area occupied by the cluster.

We did tests with circles of very different radii 
searching a reasonable tradeoff between cleanness 
(absence of a significant amount of cluster members) 
and signal-to-noise ratio
(working area not too small, to avoid small number statistics). 
The kernel density estimator technique (\citeauthor{Hand} \citeyear{Hand}) was 
applied in the VPD to these data. Details of the procedure
can be found in \cite{gala2}.
Circular Gaussian kernel functions were used, with 
Gaussian dispersion (also called smoothing parameter) $h$
elected according to 
Silverman's rule (\citeauthor{Sil} \citeyear{Sil}).
This way, the empirical frequency functions were computed
for a grid with cell size of 0.2 mas~yr$^{-1}$, well below the proper 
motion errors.
The procedure was tested for several subsamples applying different 
proper motion cutoffs and the adopted one, as in the cases of NGC~1817 and
NGC~2548, is of $|\mu| \leq$ 15 mas~yr$^{-1}$.

We finally find that the area outside a circle with a radius of 
35$\arcmin$ centred on the cluster yields a clean frequency 
function with low cluster contamination and low noise.
We can scale this field frequency function to represent
the field frequency function in the inner circle, $\Psi_{\rm f}$, 
by simply applying a factor linked to the area. The cluster
empirical frequency function can then be determined as
$\Psi_{\rm c}$~= $\Psi_{\rm c+f} - \Psi_{\rm f}$. 
Normalizing the frequency functions to volume unity we get the
probability density functions for the mixed
population (inside the circle), for the field
(outside the circle) and for the cluster (non-field)
population. These PDFs are displayed in 
Figure~\ref{nprob67}.  
We estimate the typical noise level, $\gamma$, present in the result
and we restricted the
probability calculations to the stars with cluster PDF $\geq$ 3$\gamma$.

The maximum of the cluster PDF is located at 
($\mu_{x}, \mu_{y}$)~= ($-$0.6$\pm$0.2,0.4$\pm$0.2) mas~yr$^{-1}$, 
which coincides
well with the values obtained in the parametric method.
If we transform this value to the absolute system, we get an absolute 
proper motion for the cluster of
($\mu_{\alpha}\cos\delta, \mu_{\delta}$)~= ($-$7.1$\pm$0.7,$-$7.7$\pm$0.5) mas~yr$^{-1}$.
Thirty six out of the 50 Tycho-2 stars are classified as members giving 
a mean value of
($\mu_{\alpha}\cos\delta, \mu_{\delta}$)~= ($-$8.4$\pm$2.1,$-$6.1$\pm$2.2) 
mas~yr$^{-1}$. This is fully compatible with our result, but we consider 
our figures more reliable, since they are derived from the whole sample
of stars.

\subsection{Results and discussion}

The non-parametric approach does not take 
into account the errors of the individual proper motions. 
But the full width at half maximum (FWHM) of the PDF of the 
cluster gives an estimation of the width of the distribution. 
We obtained a FWHM of $\sim$4.1$\pm$0.2~mas~yr$^{-1}$.
Taking into account the Gaussian dispersion owed to the smoothing 
parameter $h~=$ 1.34~mas~yr$^{-1}$, this would correspond to a 
r.m.s. error on proper motions of 1.55~mas~yr$^{-1}$. 
But from Zhao et al. (\citeyear{zhati}) we know that the mean proper motion 
precision is 1.24~mas~yr$^{-1}$,
what gives us an intrinsic dispersion component of 0.93~mas~yr$^{-1}$,
of the same order that the value obtained by the 
parametric membership determination ($\sigma_{c}$~= 0.89$\pm$0.05 
mas~yr$^{-1}$, $\sim$ 4~km~s$^{-1}$ at the distance of 900~pc).

\begin{figure}
\begin{center}
\resizebox{7cm}{!}{\includegraphics{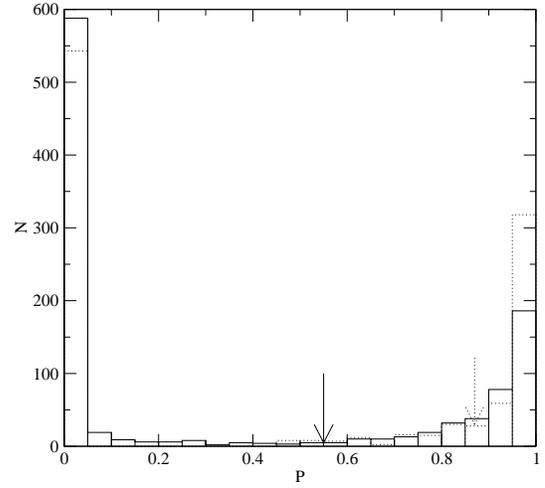}}
\end{center}
\caption{The histogram of cluster membership probability 
of M~67. The solid line gives the results for classical parametric 
method $P_{\rm P}$ (Section \ref{parametric}), 
while the dotted line corresponds to the
non-parametric approach $P_{\rm NP}$ (Section \ref{nonparametric}). 
The arrows mark the limiting 
probabilities for member selection for each method.}
\label{fig767}
\end{figure}

Membership probabilities are computed according to Eq.
\ref{probabilitat}, both from the classical, parametric
frequency functions ($P_{\rm P}$) and from the non-parametric, empirical 
frequency funcions ($P_{\rm NP}$).
The cluster membership probability histogram (Figure~\ref{fig767}) 
shows a clear separation between cluster members and field stars
in both approaches (the solid line being the classical parametric method,
dotted line the non-parametric approach).
The non-parametric technique yields
an expected number of cluster members from the integrated volume of the 
cluster frequency function in the VPD areas of high cluster density, 
where PDF$\geq$ 3$\gamma$. 
The expected amount of cluster members in this VPD area is
of 393. Sorting the sample in order of decreasing 
non-parametric membership probability, $P_{\rm NP}$, the first 393 stars 
are the most probable cluster members. The minimum value of the non-parametric 
probability (for the 393-rd star) is $P_{\rm NP}~= 0.87$. 

To decide where to set the 
limit among members and non-members in the list sorted in order of 
decreasing parametric membership probability, $P_{\rm P}$, we accept 
the size of the cluster predicted by the non-parametric method, 393 
stars. Thus we consider that the 393 stars of highest $P_{\rm P}$ are the most 
probable members, according to the results of the parametric 
technique. The minimum value of the parametric probability 
(for the 393-rd star) is $P_{\rm P}~= 0.55$. It is worth noting that, contrary
to what happened with the previous two clusters 
studied by us, and in other cases
found in the literature \citep{gala2}, in this case the original parametric
segregation \citep{zhati} at $P_{\rm P} > 0.8$ was underestimating, and not
overestimating, the number of probable members.
This is because M~67 is much more populated, and displays
a higher constrast against the field, than the other clusters.
The need to reach a $P_{\rm P}$ cutoff much lower than the $P_{\rm NP}$ cutoff 
illustrates the effects of an imperfect modelling of the wings of the real 
distribution in the parametric method. 
Also, the importance of having some objective criterium to 
decide the $P_{\rm P}$ cutoff is evident from our results.

With these limiting probabilities ($P_{\rm NP}\geq 0.87$; $P_{\rm P}\geq 0.55$), 
we get a 97$\%$ (1014 stars) of agreement  
in the segregation yielded by the two methods. 
Table~7 lists the cross-identifications with \cite{zhati} 
and \cite{Sanders77}, the $P_{\rm P}$ and the $P_{\rm NP}$ for the 1046 stars. 

\addtocounter{table}{1}

\begin{figure*}
\begin{center}
\resizebox{11cm}{!}{\includegraphics{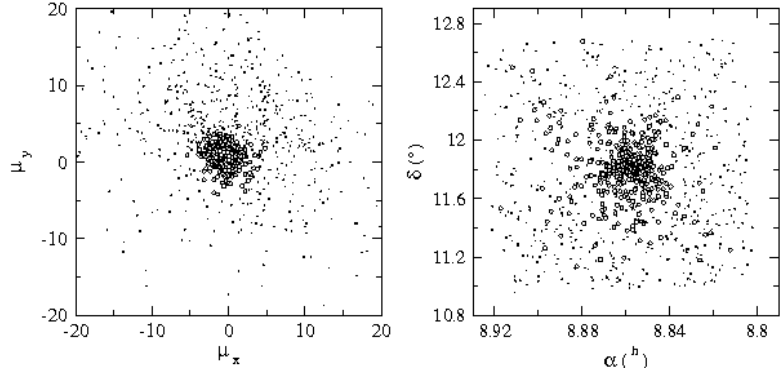}}
\caption{The proper motion vector-point diagram (VPD) and spatial distribution
 of stars in M~67 region 
(empty circles for the astrometric probable members of M~67 and
    dots for field stars). Units of relative proper motions are mas~yr$^{-1}$.}
\label{pmad67}
\end{center}
\end{figure*}

As in \cite{Bal04a}, 
we accept as probable 
members those stars classified
as member by at least one of the two methods.
This way we get a list of 412 astrometric probable member stars.
Figure~\ref{pmad67} shows the proper motion VPD
and the sky distribution for all the measured
stars, where empty circles
denote an astrometric probable member of M~67,
and all other stars are considered field stars and they are indicated
by dots.

A comparison of the cluster/field segregation for the 143 stars 
in common with the radial velocity studies 
by \cite{Mathieu, Mathieu90} is given 
in Tables \ref{vr67M} (candidate members) and \ref{vr67NM} (non-members). 
The radial velocities have typical standard deviations
for a set of measurements of any given star that range 
between 0.5 and 0.8 km~s$^{-1}$.
To quantify the differences in the segregation, we do as in 
\cite{Bal04a,Bal05} and set an agreement 
index $P_{\rm c}$ to 1 if the parametric probability, $P_{\rm P}$, agrees with the radial velocity 
segregation, 2 if the non-parametric probability, $P_{\rm NP}$, agrees, 
3 if both probabilities, $P_{\rm P}$ and $P_{\rm NP}$, agree and 0 if none does. 
We find 125 out of 143 stars with $P_{\rm c} >$~0, that is 87$\%$ agreement
with the radial velocities segregation. 
This 13$\%$ of disagreement consists of  
15 stars out of 41 being considered non-members on the basis 
of proper motions,
while only 3 out of 102 were found to be astrometric 
members but considered non-members on the basis of radial velocities.  
Those three have only one measurement of their radial velocities
and two of them are known to be spectroscopic binaries (SB), 
therefore the measurement may
be different from the systemic velocity and thus
they cannot be completely ruled out as members.  

If we compare the parametric and non-parametric methods, the behaviour 
is rather similar. For the 
parametric method we find a total of 121 stars (85$\%$) whose membership
assignation coincides with the radial velocity criterion, while
for the non-parametric method this amounts to 124 stars (87$\%$).

\begin{table*}
\caption {The cross-identification of stars considered as candidate members by us, 
with the radial velocities sudies by \cite{Mathieu, Mathieu90} and
the comparison of those with the parametric ($P_{\rm P}$) and 
non-parametric ($P_{\rm NP}$) probabilities. 
See text for explanation of the agreement index $P_{\rm c}$. }

{\scriptsize
\begin{center} 
\begin {tabular} {cccccccccccc}
\hline
ID$_{\rm BDA}$ & ID$_{\rm Zhao}$ & $P_{\rm P}$ & $P_{\rm NP}$ & $P_{\rm c}$ & $Vr^*$ & 
ID$_{\rm BDA}$ & ID$_{\rm Zhao}$ & $P_{\rm P}$ & $P_{\rm NP}$ & $P_{\rm c}$ & $Vr^*$  \\
\hline
4 & 818 & 0.94 & 0.96 & 3 & 34.0 $\pm$2.0 &        185 & 958 & 0.79 & 0.94 & 3 & 33.4 $\pm$0.0  \\
  16 & 829 & 0.88 & 0.96 & 3 & 33.2 $\pm$0.9 &    190 & 989 & 0.80 & 0.94 & 0 & 43.6 $\pm$0.0 SB \\ 
18 & 521 & 0.92 & 0.95 & 3 & 32.9 $\pm$0.5 &       192 & 603 & 0.98 & 0.97 & 3 & 32.7 $\pm$1.1 \\ 
  20 & 834 & 0.96 & 0.97 & 3 & 33.1 $\pm$1.0 &    193 & 999 & 0.98 & 0.97 & 3 & 33.7 $\pm$0.5 \\    
22 & 839 & 0.95 & 0.96 & 3 & 34.4 $\pm$1.9 SB1 &   195 &1017 & 0.97 & 0.98 & 3 & 33.87$\pm$0.12 SB1\\
  28 & 840 & 0.97 & 0.97 & 3 & 33.7 $\pm$1.0 &    210 &1005 & 0.98 & 0.97 & 3 & 33.9 $\pm$1.1 \\    
30 & 845 & 0.97 & 0.97 & 3 & 33.6 $\pm$0.4 &       215 & 608 & 0.98 & 0.98 & 3 & 33.6 $\pm$0.6 \\
  37 & 846 & 0.97 & 0.97 & 3 & 33.7 $\pm$0.3 &    216 &1010 & 0.97 & 0.98 & 3 & 33.03$\pm$0.13 SB1 \\
46 & 843 & 0.98 & 0.97 & 3 & 33.8 $\pm$0.1 &       217 &1002 & 0.97 & 0.97 & 3 & 33.3 $\pm$0.4 \\
  48 & 848 & 0.96 & 0.98 & 3 & 33.0 $\pm$0.4 &    218 &1006 & 0.98 & 0.97 & 3 & 34.0 $\pm$0.5 \\    
51 & 849 & 0.96 & 0.97 & 3 & 34.6 $\pm$0.9 &       219 &1007 & 0.98 & 0.98 & 3 & 33.4 $\pm$0.3 SB1\\
  54 & 863 & 0.95 & 0.97 & 3 & 34.5 $\pm$0.7 &    223 & 609 & 0.68 & 0.94 & 3 & 32.8 $\pm$0.4 \\    
55 & 862 & 0.75 & 0.92 & 0 & 42.1 $\pm$0.0 SB &    224 &1011 & 0.95 & 0.96 & 3 & 32.55$\pm$0.07 SB1\\
  72 & 876 & 0.97 & 0.98 & 3 & 33.3 $\pm$0.6 &    226 &1013 & 0.98 & 0.98 & 3 & 32.5 $\pm$1.0 \\    
79 & 860 & 0.97 & 0.98 & 3 & 34.1 $\pm$1.7 &       227 &1004 & 0.98 & 0.97 & 3 & 32.8 $\pm$0.7 \\
  84 & 966 & 0.78 & 0.94 & 3 & 34.1 $\pm$0.4 &    231 &1029 & 0.97 & 0.97 & 3 & 32.5 $\pm$0.6 \\    
86 & 933 & 0.81 & 0.86 & 1 & 27.6 $\pm$0.0 &       236 &1025 & 0.98 & 0.98 & 3 & 33.82$\pm$0.18 SB1\\
  88 & 927 & 0.96 & 0.97 & 3 & 34.13$\pm$0.23 SB2&237 & 611 & 0.97 & 0.98 & 3 & 34.7 $\pm$0.7 \\    
95 & 886 & 0.98 & 0.98 & 3 & 34.7 $\pm$2.4 &       238 &1028 & 0.75 & 0.94 & 3 & 38.3 $\pm$0.0 \\
  96 & 929 & 0.96 & 0.98 & 3 & 33.1 $\pm$0.7 &    241 &1027 & 0.96 & 0.96 & 3 & 33.2 $\pm$0.7 \\    
101 & 930 & 0.97 & 0.98 & 3 & 32.9 $\pm$0.9  &     243 &1026 & 0.96 & 0.96 & 3 & 33.4 $\pm$0.5 \\
 102 & 893 & 0.98 & 0.97 & 3 & 36.17$\pm$0.17 SB1&244 &1032 & 0.93 & 0.96 & 3 & 33.55$\pm$0.05 SB1 \\
104 & 931 & 0.95 & 0.97 & 3 & 33.5 $\pm$0.4  &     248 & 612 & 0.98 & 0.98 & 3 & 34.0 $\pm$1.3 \\
 105 & 883 & 0.80 & 0.94 & 3 & 34.3 $\pm$0.7 &    252 & 618 & 0.97 & 0.96 & 3 & 32.5 $\pm$1.0 \\    
108 & 894 & 0.51 & 0.93 & 2 & 34.7 $\pm$0.6  &     255 &1043 & 0.96 & 0.97 & 3 & 31.3 $\pm$0.3 \\
 111 & 891 & 0.98 & 0.97 & 3 & 33.70$\pm$0.22 SB1&256 & 226 & 0.96 & 0.98 & 3 & 33.2 $\pm$0.9 \\    
115 & 918 & 0.97 & 0.98 & 3 & 34.4 $\pm$0.5  &     262 &1046 & 0.97 & 0.97 & 3 & 31.6 $\pm$0.7 \\
 117 & 887 & 0.98 & 0.98 & 3 & 34.4 $\pm$0.3 SB2 &266 & 647 & 0.78 & 0.94 & 3 & 34.3 $\pm$0.4 \\    
119 & 924 & 0.97 & 0.97 & 3 & 34.34$\pm$0.21 SB2 & 271 &1058 & 0.93 & 0.96 & 3 & 34.1 $\pm$0.8 \\
 124 & 912 & 0.97 & 0.98 & 3 & 29.6 $\pm$0.7 SB & 272 & 645 & 0.97 & 0.97 & 3 & 31.8 $\pm$0.4 \\    
127 & 911 & 0.97 & 0.98 & 3 & 33.3 $\pm$0.8  &     281 &1060 & 0.96 & 0.97 & 3 & 33.4 $\pm$0.8 \\
 130 & 910 & 0.94 & 0.98 & 3 & 33.7 $\pm$0.7 &    286 & 697 & 0.35 & 0.92 & 2 & 33.6 $\pm$0.5 \\    
131 & 972 & 0.95 & 0.96 & 3 & 33.3 $\pm$2.0 SB2 &  287 & 698 & 0.98 & 0.97 & 3 & 32.8 $\pm$0.5 \\
 134 & 907 & 0.98 & 0.97 & 3 & 32.5 $\pm$1.2 SB1& 289 & 683 & 0.94 & 0.96 & 3 & 32.7 $\pm$0.4 \\    
135 & 908 & 0.97 & 0.97 & 3 & 34.3 $\pm$0.6  &     291 & 700 & 0.96 & 0.97 & 3 & 34.1 $\pm$0.1 \\
 136 & 974 & 0.94 & 0.96 & 3 & 32.87$\pm$0.12 SB1&305 & 133 & 0.52 & 0.92 & 2 & 34.2 $\pm$0.9 \\    
141 & 947 & 0.86 & 0.95 & 3 & 33.6 $\pm$0.4  &     1182 & 174 & 0.95 & 0.98 & 3 & 34.6 $\pm$0.9 \\
 143 & 937 & 0.91 & 0.96 & 3 & 32.93$\pm$0.07 SB1&2079& 172 & 0.98 & 0.98 & 3 & 34.4 $\pm$0.5 \\    
149 & 949 & 0.98 & 0.97 & 3 & 35.5 $\pm$0.8  &     2087 & 211 & 0.94 & 0.97 & 3 & 31.5 $\pm$3.0\\
 151 & 971 & 0.81 & 0.94 & 3 & 33.9 $\pm$0.5 &    3035 &1001 & 0.98 & 0.98 & 3 & 34.1 $\pm$0.5 \\   
157 & 976 & 0.98 & 0.98 & 3 & 33.6 $\pm$0.5  &     3116 & 642 & 0.98 & 0.97 & 3 & 33.60$\pm$0.13 SB1\\
 163 & 995 & 0.98 & 0.98 & 3 & 33.8 $\pm$0.8 &    4004 & 939 & 0.98 & 0.98 & 3 & 33.48$\pm$0.19 SB2\\
164 & 983 & 0.79 & 0.94 & 3 & 33.3 $\pm$0.4  &     4096 & 578 & 0.97 & 0.97 & 3 & 34.0 $\pm$0.8 \\
 166 &   4 & 0.97 & 0.98 & 3 & 33.3 $\pm$0.3 &    7591 & 135 & 0.91 & 0.96 & 3 & 33.4 $\pm$0.8  \\ 
170 & 953 & 0.60 & 0.93 & 3 & 33.59$\pm$0.10 SB1 & 7657 & 505 & 0.28 & 0.87 & 2 & 33.2 $\pm$0.9 \\
 174 & 961 & 0.97 & 0.98 & 3 & 36.2 $\pm$4.5&     8402 & 231 & 0.83 & 0.94 & 3 & 33.6 $\pm$0.4 \\   
176 & 914 & 0.97 & 0.98 & 3 & 32.5 $\pm$0.4 SB1 &  8524 & 639 & 0.69 & 0.89 & 3 & 33.6 $\pm$0.7 \\
 180 & 985 & 0.98 & 0.97 & 3 & 35.0 $\pm$0.8 &    8571 & 253 & 0.94 & 0.95 & 3 & 34.2 $\pm$0.4 \\   
181 & 955 & 0.98 & 0.98 & 3 & 33.3 $\pm$0.7 &      8792 & 282 & 0.85 & 0.93 & 3 & 33.1 $\pm$0.5 \\
 182 & 956 & 0.97 & 0.98 & 3 & 32.2 $\pm$2.2 &    8808 & 277 & 0.85 & 0.94 & 3 & 29.6 $\pm$1.0 \\   
184 & 987 & 0.98 & 0.97 & 0 & 61.4 $\pm$0.0&       8832 & 713 & 0.97 & 0.97 & 3 & 33.5 $\pm$0.7 \\
\hline
\end{tabular}
\end{center} 
\label{vr67M}
(*) Zero errors mean that only one measurement was taken and no dispersion 
could be calculated. SB1: Single-lined spectroscopic binary. 
SB2: Double-lined spectroscopic binary.
}
\end{table*}

\begin{table*}
\caption {The same as Table \ref{vr67M}, for stars considered as non-members by us.
}

{\scriptsize
\begin{center} 
\begin {tabular} {cccccccccccc}
\hline
ID$_{\rm BDA}$ & ID$_{\rm Zhao}$ & $P_{\rm P}$ & $P_{\rm NP}$ & $P_{\rm c}$ & $Vr^*$ & 
ID$_{\rm BDA}$ & ID$_{\rm Zhao}$ & $P_{\rm P}$ & $P_{\rm NP}$ & $P_{\rm c}$ & $Vr^*$  \\
\hline
8 & 822 & 0.00 & 0.70 & 3 &  29.1 $\pm$1.8&        7116 &  27 & 0.00 & 0.00 & 3 &  13.7 $\pm$0.0 \\   
 10 & 136 & 0.00 & 0.00 & 3 &  16.3 $\pm$0.0 &      7232 &  52 & 0.00 & 0.00 & 3 &   4.2 $\pm$0.0 \\
45 & 857 & 0.00 & 0.00 & 3 & -14.0 $\pm$0.0 &      7251 & 434 & 0.00 & 0.47 & 0 &  34.56$\pm$0.14 SB1\\
 49 &  3  & 0.00 & 0.00 & 3 &  28.1 $\pm$0.0 &      7335 &  75 & 0.00 & 0.00 & 3 & -18.9 $\pm$0.0 \\
144 & 951 & 0.00 & 0.00 & 0 &  33.72$\pm$0.14 SB1& 7350 &  72 & 0.00 & 0.00 & 3 & -13.0 $\pm$0.0 \\   
 155 & 943 & 0.00 & 0.00 & 3 &   3.6 $\pm$1.0 &     7378 & 458 & 0.00 & 0.00 & 3 &  57.8 $\pm$0.0 \\
173 & 963 & 0.00 & 0.00 & 0 &  33.37$\pm$0.20 SB1& 7434 &  94 & 0.00 & 0.65 & 0 &  33.7 $\pm$0.9 \\   
 200 & 601 & 0.00 & 0.00 & 3 &  12.0 $\pm$0.4&      7440 &  98 & 0.00 & 0.00 & 3 &  14.95$\pm$0.11 SB1\\
206 & 602 & 0.00 & 0.00 & 0 &  34.8 $\pm$0.3 &     7445 & 85  & 0.00 & 0.72 & 0 &  33.8 $\pm$1.0 \\   
 240 & 215 & 0.17 & 0.84 & 3 &  43.3 $\pm$0.15 SB1&7488 & 485 & 0.06 & 0.84 & 0 &  33.0 $\pm$0.3 \\   
242 &1024 & 0.00 & 0.00 & 3 &  -1.1 $\pm$0.4 &      7515 & 489 & 0.06 & 0.84 & 3 &  27.6 $\pm$0.6 \\
 277 & 648 & 0.00 & 0.00 & 3 &   1.6 $\pm$0.0 &    7551 & 104 & 0.00 & 0.67 & 3 &   8.9 $\pm$0.0 \\   
2086 & 210 & 0.00 & 0.00 & 0 &  33.3 $\pm$0.6 &     7663 & 492 & 0.03 & 0.00 & 3 &   1.5 $\pm$0.0 \\
 3128 & 620 & 0.00 & 0.00 & 3 & -20.7 $\pm$0.0 &   7674 & 496 & 0.00 & 0.00 & 0 &  33.7 $\pm$0.5 \\   
4166 & 481 & 0.00 & 0.00 & 3 &  -2.5 $\pm$0.0 &     8135 & 566 & 0.00 & 0.78 & 0 &  34.2 $\pm$0.4 \\
 4168 & 482 & 0.00 & 0.00 & 3 &  39.5 $\pm$0.4 &   8355 & 623 & 0.00 & 0.63 & 3 &  49.3 $\pm$0.0 \\   
 6469 & 453 & 0.00 & 0.00 & 0 &  34.1 $\pm$0.2 &    8522 & 657 & 0.00 & 0.00 & 3 &  12.6 $\pm$0.0 \\
6470 & 468 & 0.00 & 0.70 & 0 &  33.3 $\pm$0.3 &    8533 & 632 & 0.00 & 0.76 & 3 &  67.8 $\pm$0.5 \\   
 6474 & 495 & 0.00 & 0.73 & 3 &   9.5 $\pm$0.6&     8557 & 250 & 0.00 & 0.00 & 0 &  33.9 $\pm$0.5\\
6514 & 261 & 0.00 & 0.00 & 0 &  34.5 $\pm$0.7 &    9015 & 744 & 0.10 & 0.81 & 0 &  33.5 $\pm$1.6 \\   
 7112 & 37  & 0.00 & 0.00 & 3 &  -9.0 $\pm$0.0 &      &       &     &      &   &   \\
\hline
\end{tabular}
\end{center} 
(*) Zero errors mean that only one measurement was taken and no dispersion 
could be calculated. SB1: Single-lined spectroscopic binary. 
SB2: Double-lined spectroscopic binary.
}
\label{vr67NM}
\end{table*}

\subsection{Magnitude effects in UCAC2 and proper motions as indicator of binarity}

\begin{figure}
\resizebox{\hsize}{!}{\includegraphics{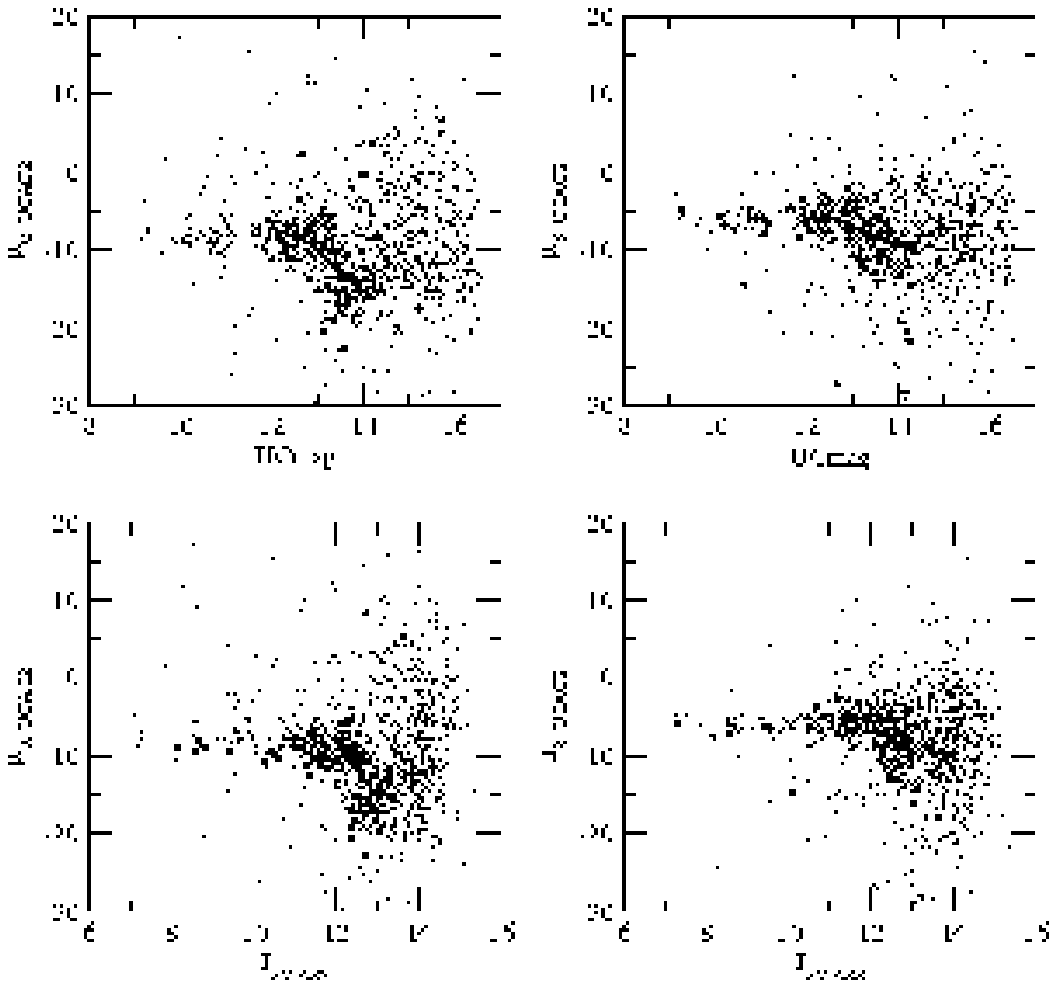}}
\caption{UCAC2 proper motion vs\ magnitude 
for M~67 area. Top row: $\mu_\alpha$ and $\mu_\delta$ vs\ the UCAC2 
magnitude \citep{UCAC2}; bottom row: vs\ 2MASS $J$ magnitude \citep{2MASS}. 
Proper-motion 
errors are about 1--3 mas~yr$^{-1}$ for stars to 12th magnitude, and about 4--7
mas~yr$^{-1}$ for fainter stars up to 16th magnitude.
UCAC2 proper motions are taken from Tycho-2 \citep{tyc2} catalogue down
to its magnitude limit $V\sim$12. From this magnitude, and down to
the limiting magnitude of our selection of candidate members (marked 
as empty circles),
there is a systematic trend depicted with a thick line
}
\label{UCACmu}
\end{figure}

The internal dynamics of non-resolved multiple 
systems could induce a motion of 
the measured photocentre different from the motion of the barycentre. 
This effect 
has to be null for equal-mass binaries and almost undetectable
for binaries with components of extremely 
different masses, and should reach a maximum somewhere in between. 
When looking to 
the histogram of proper motion modules, this effect would skew 
the distribution inducing a longer tail towards high values.

\cite{Bica05} have made an attempt of 
identifying high-velocity stars as unresolved binary cluster members and, as a
result, for mapping and quantifying the binary component in 
colour-magnitude diagrams.
They analysed 9 open clusters using UCAC2 proper motions \citep{UCAC2} 
and 2MASS 
photometry \citep{2MASS}. As a test case of the method, they use M~67. 
Their study of the modulus of the proper motion in the cluster area yields a double-peaked
distribution instead of the skewed Gaussian that, in principle, 
could be expected. 
The authors associate the peaks to 
high-velocity and low-velocity populations in 
the cluster, and the high-velocity group is interpreted as produced by 
unresolved binary systems. This behaviour is found in the core and in the halo 
of the cluster. The authors claim internal dispersions of the order of 
$\sim$6~km~s$^{-1}$ for single stars and around $\sim$11~km~s$^{-1}$
for unresolved binaries, and a difference between the two peaks
of more than 20$-$30~km~s$^{-1}$. These strikingly unrealistic numbers and the
lack of any error estimation adds suspicion to the fact that 
all the stars from the high-velocity peak are 
dimmer than $J \sim$ 12 (Fig.~6 in their paper), when the abundance 
of binaries among the bright members of M~67 is well known. 

A similar study using the proper motions by \cite{zhati} gives as a result a
bell-shaped distribution with a tail towards higher velocities. We
were not able to find a double-peaked distribution in the core nor in 
the halo. Different tries of dividing the distribution did not show any
relation to binarity when checked  
on the corresponding colour-magnitude diagram.

As it is well depicted by the study of \cite{Platais03}, the presence of 
residual colour/magnitude terms in any proper motion analysis is almost 
unavoidable. This is specially true when one deals with an all-sky catalogue,
where the proper motions are derived from different first epoch sources. Aware
of these facts, we have checked for magnitude terms in UCAC2 \citep{UCAC2} proper motions.
We plot proper motion vs\ magnitude in Fig.~\ref{UCACmu}. 
The UCmag magnitude in the top row is in the UCAC bandpass 
(579-642~nm, between $V$ and $R$) and should be considered approximate. 
UCAC2 proper motions are derived by using over 140 ground- and space-based 
catalogues, including Tycho-2 \citep{tyc2} and the AC2000.2 \citep{AC2000}.
In the bottom row, we
have plotted the 2MASS $J$ magnitude as a double check of the results.
Cluster member candidates from our selection 
are marked as empty circles. We can note how 
the UCAC2 proper motions of these stars
show a trend with magnitude for magnitudes fainter than from $V\sim 12$, 
highlighted by the regression fit marked by a 
thick line. 
The behaviour detected by \cite{Bica05} is not related to binarity, but due 
to systematics in the proper motion data. 
The use of proper motion data from UCAC2 Catalogue should be carefully
taken to avoid systematics.


\section{Colour-Magnitude Diagrams}

\begin{figure*}
\resizebox{\hsize}{!}{\includegraphics{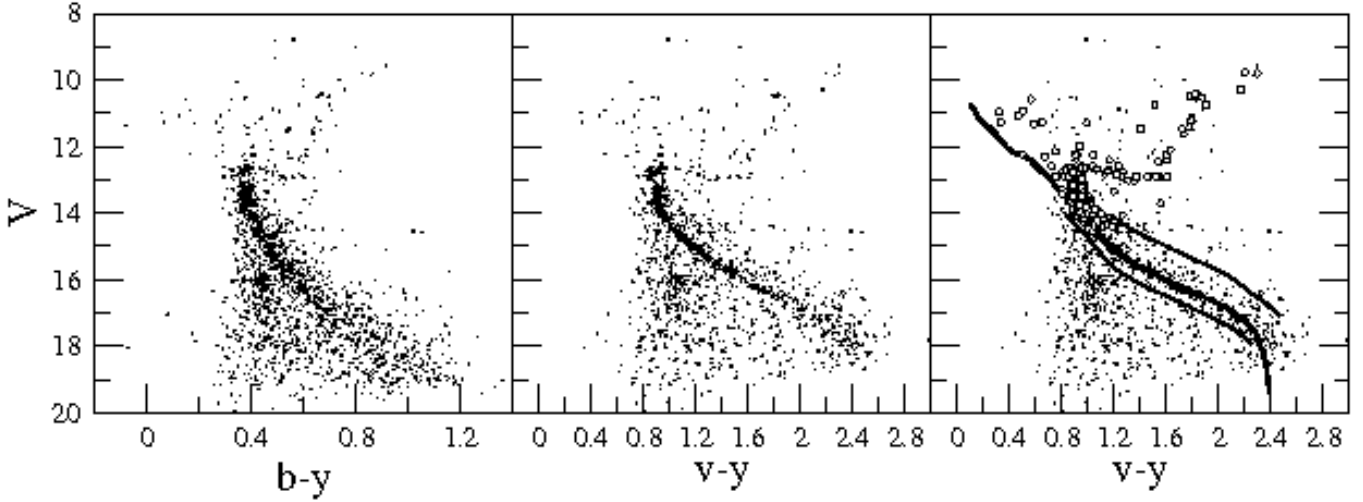}}
\caption{Colour-magnitude diagrams of the M~67 area.
Figure on the right shows the astrometric probable members as empty circles. 
Thick line is a shifted ZAMS, with the chosen margin for candidate members
($V+0.5$,$V-1$) in thin lines. See text for details.}
\label{HR67}
\end{figure*}

	We use the $V$ vs\ $(v-y)$ colour-magnitude diagram 
for our study (Fig.~\ref{HR67} right and centre) because it 
defines the main-sequence of a cluster significantly better 
than the traditional $V$ vs\ $(b-y)$ diagram \citep{Mei00} 
in presence of extinction. 
The colour-magnitude diagram of all the stars in the area displays a
fairly well defined main sequence. The advanced age of this cluster is obvious
at first sight looking at the diagram. Moreover, the sequence of binaries can
be easily followed.

\begin{figure}
\resizebox{\hsize}{!}{\includegraphics{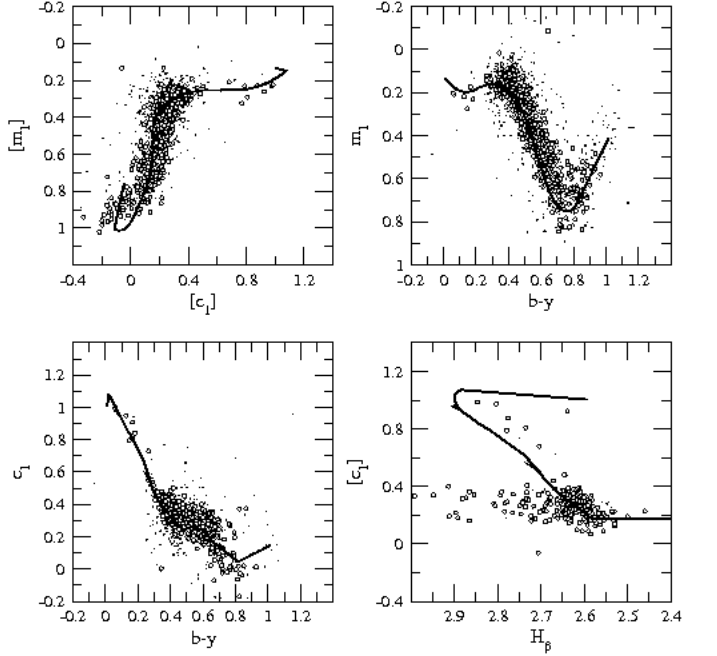}}
\caption{The colour-colour diagrams of M~67.
           Empty circles denote candidate members of M~67, chosen with
astrometric and non-astrometric criteria as explained in Section~4.1. 
           The thick line is the solar ZAMS standard relation shifted 
by $E(b-y)$ = 0.03 when necessary. Evolutionary and binarity effects are 
noticeable in these diagrams.
Some of the candidate members, known to be multiple stars,
have very discordant values.} 
\label{colour67}
\end{figure}

\subsection{Selection of candidate member stars}

Unfortunately, proper motions are only available for the 
brightest stars in the area. Photometric 
measurements help to reduce the possible field contamination in the
proper motion membership ---among bright stars---,
as well as to enlarge the selection of members towards faint magnitudes. 
Our astrometric segregation of member stars has a limiting magnitude 
of $V \sim$ 15.5. 
From this magnitude down to $V\sim 18$ we construct a ridge line
following a fitting of the   
observational zero-age main sequence (ZAMS) 
\citep{Craw75, Craw78, Craw79, Hil83, Ols84},  
on the $V$ vs\ $(v-y)$ diagram. 
A selection of stars based on the distance to this ridge line is then
performed. 
The width of the sequence
defined by astrometric candidate members allows us to establish the 
margins for the selection of fainter stars. 
The chosen margin includes all the stars 
between $V+0.5$ and $V-1$ from the ridge line, as shown in the
right panel of Fig.~\ref{HR67}.
This magnitude range allows for photometric errors in the two axis, 
intrinsic dispersion around the ZAMS and multiplicity effects. 
The interval of 1 magnitude at the bright side comprises
the width of the sequence and is large enough to allow for equal-mass
binaries ($\Delta V = 0.75$) and for most triples ($\Delta V = 1.19$
for equal-mass systems). The interval of 0.5 magnitudes at the faint side
agrees with the observed width of the main sequence and is larger
than the expected photometric errors. 
Luckily, the high density of M~67 and its outstanding contrast against the 
population of field stars (shown by the comparison of the density functions 
in Fig.~\ref{nprob67}) makes the fine-tuning of these margins 
to be non-critical: the number of non-members introduced by our choice
is low, and the loss of true members is minimised.  

Extreme outliers have been rejected making use of the colour-colour
diagrams (Fig.~\ref{colour67}) with the help of the standard relations\footnote 
{The asymmetry of the candidate members in the $m_1$ vs $b-y$ diagram 
is due to: binarity, on the one hand, which in the case of the cool stars 
it always mimics a metal deficiency; and on the other hand, 
to the presence of giants, which are located towards 
lower $m_1$ than for dwarfs for the same $b-y$ \citep{Ols84}.}  
\citep{Craw75, Craw78, Craw79, Hil83, Ols84}.  
The astrometric and photometric selection gives 
a final set of
776 candidate members. They are plotted 
in Fig.~\ref{colour67}
as filled circles in the $[m_1] - [c_1]$, $m_1 - (b-y)$, $c_1 - (b-y)$ and 
$[c_1] - H_{\beta}$ diagrams. 
This photometric selection is known to contain a certain amount of 
field contamination, reason why we prefer to talk about ``candidate members'', 
rather than ``bona fide members''. Anyhow, the selection done leads to a set of 
stars overwhelmingly dominated by true members: 
the methods applied in Sect.~5 to 
derive the physical parameters of the cluster are readily capable of extracting
reliable mean values from such a sample.

\section{Physical parameters of the cluster}

The stars of the area selected as candidate cluster members were
classified into photometric regions and its physical parameters determined, 
following the algorithm described in Masana et al.\ (\citeyear{Mas}) and  
\cite{Jordi97}. The algorithm uses 
$uvby-H_{\beta}$ photometry and standard relations among colour indices
for each of the photometric regions of the HR diagram. 

\subsection{Distance, reddening and metallicity}

\begin{figure}
\resizebox{\hsize}{!}{\includegraphics{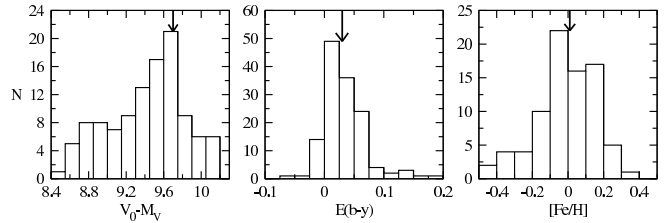}}
\caption{The histograms of the distance modulus, reddening and 
metallicity of the selected member stars of M~67 with $H_{\beta}$
measurements. The arrows indicate the mean values adopted for the cluster.}
\label{pf67}
\end{figure}

Only 251 stars among the 776 candidate members have $H_{\beta}$
measurements. So the computation of physical parameters is only 
possible for that subset.   
        The results are shown in Fig.~\ref{pf67}. 
Excluding peculiar stars and those with inconsistence among their 
photometric indices and applying an average with a  
2$\sigma$ clipping to that subset, 
we found a mean
reddening value of $E(b-y)$~=~0.03$\pm$0.03
(corresponding to $E(B-V)$~=~0.04)
and a mean distance modulus of $V_0-M_V$~=~9.4$\pm$0.4.
The distance modulus may be biased towards short values
due to the presence of multiple stars treated as if they
were single ones.
A substantial fraction of binaries tend to have equal mass companions.
These binaries would be well-separated from the cluster main-sequence, so 
they can be easily detected and removed. However, low mas ratio binaries would
remain, and might systematically reduce the estimated distance since they are 
brighter and redder than single cluster members. 
If we set a conservative limit for stars with lower 
value than 9.4 in the distance modulus, 
we can see from the $V$ vs\ $b-y$ diagram that these lower values 
mainly correspond to stars above the main sequence of the cluster 
which are most probably multiple stars. Excluding those we can get 
a value of $V_0-M_V$~=~9.7$\pm$0.2 from the study of 66 stars that 
can be confidently considered singles. 
 
Metallicity is better calculated studying only the 81 F and G type stars in 
our sample following \cite{masana94}. 
We find a mean value of [Fe/H]~=~0.01$\pm$0.14. 

\subsection{Age}

\begin{figure}
\resizebox{\hsize}{!}{\includegraphics{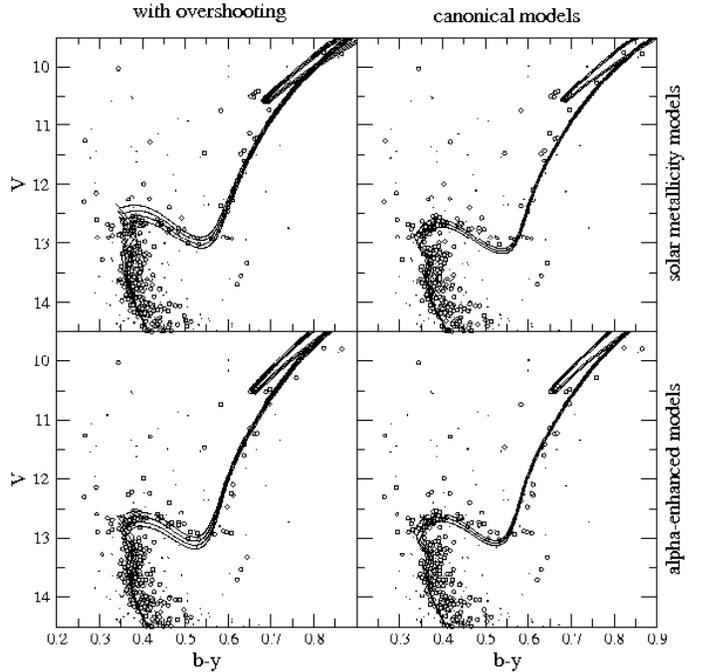}}
\caption{Isochrones from \cite{Pie} for scaled solar models (top row) of
solar metallicity and ages   
of 4.0, 4.2, 4.4 and 4.6~Gyr for models with overshooting
(left panel) and ages of 3.4, 3.6, 3.8~Gyr for canonical 
models (right panel). 
Isochrones from \cite{Pie06} for alpha-enhanced models (bottom row) 
of solar metallicity and ages 
of 3.6, 3.8, 4.0 and 4.2~Gyr for models with overshooting
(left panel) and ages of 3.0, 3.2, 3.4~Gyr for canonical 
models (right panel). 
}
\label{iso67}
\end{figure}

	The publication by Clem et al.\ (\citeyear{Clem}) 
of empirically constrained colour-temperature relations in the Str\"omgren
system makes possible an isochrone fitting to our results. The least 
unsatisfactory fitting is
found for the \cite{Pie,Pie06} tracks. Figure~\ref{iso67}
shows the isochrones 
from \cite{Pie} for scaled solar models 
of solar metallicity and ages   
of 4.0, 4.2, 4.4 and 4.6~Gyr for models with overshooting
and of 3.4, 3.6, 3.8~Gyr for canonical models.
The addopted reddening and distance modulus are $E(b-y)$~= 0.03 
and $V_0 - M_V$~= 9.7; 
from \cite{Pie06} for $\alpha$-enhanced models 
of solar metallicity and ages 
of 3.6, 3.8, 4.0 and 4.2~Gyr for models with overshooting
and of 3.0, 3.2, 3.4~Gyr for canonical models.
For this case, the addopted reddening and distance modulus are $E(b-y)$~= 0.05 
and $V_0 - M_V$~= 9.7. Empty circles are candidate members.
Canonical models give lower ages. 
$\alpha$-enhanced models reproduce better the behaviour of the red giants 
clump and red giant branch, but seem to behave worse in the lower giants branch and 
subgiants branch where models with overshooting seem to give a better fit.
It seems apparent that none of the models provides a good fit over all 
the areas of the colour-magnitude diagram.
We decided to adopte a compromise between these isochrones and give an 
estimation of the age of $t$ = 4.2$\pm$0.2~Gyr ($\log t$~= 9.62$\pm$0.02)
in agreement with previous estimates by other outhors.

\section{Multiple Star Systems, Blue Stragglers and other peculiars}

\subsection{Multiple Star Systems}

	The high binary content of M~67 was already noted by \cite{Racine}:
``more than half of the M~67 main-sequence stars between G2 and K5 appear
to be unresolved multiple stars".
\cite{Mont} studied the distribution of binaries and calculated a 22$\%$ of
equal-mass component binaries. Trying to account for the binaries with
low mass ratios when studying the distribution of multiple stars from a fiducial
main sequence, they found a ratio of 38$\%$. Due to the possible presence
of very low mass ratio binaries, this percentage is a lower limit.
\cite{Fan} reconsidered the mass ratios
and number of binaries in M~67, and using models leaving the mass ratio 
to randomly vary from zero to one, they found that the
true binary fraction in this cluster depends critically on how to account for
the contribution of low mass-ratio binaries. From the models, around a 50$\%$ 
of binaries seems a plausible scenario, with a binary mass ratio distribution 
more consistent with being random than double-peaked.

\begin{figure}
\begin{center}
\resizebox{9cm}{!}{\includegraphics{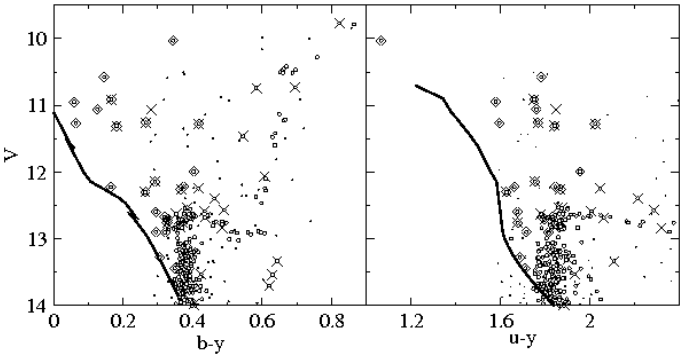}}
\caption{Multiple star systems (marked as crosses) and blue stragglers 
(diamonds) in a colour magnitude diagram in $b-y$ (left panel) 
and $u-y$ (right panel) of M~67. Empty circles are candidate member stars. 
Dots are non-member stars. A ZAMS is shown as reference.
}
\label{SBBS67}
\end{center}
\end{figure}

\begin{table*}
\caption {The cross-identification of known multiple star systems 
and their astrometric segregation from this work (last column). First column is our
identification number (Table~3), second column that from BDA 
and third that from \cite{Sanders77}.} 

{\scriptsize
\begin{center} 
\begin {tabular} {ccccccccc}
\hline
ID$_{\rm 3}$ & ID$_{\rm BDA}$  & ID$_{\rm S}$ & $b-y$ & $V$ & $m_1$ & $c_1$ & $H_{\beta}$ & M/NM  \\ 
\hline
  609&   22&  821&  0.371$\pm$0.001& 12.764$\pm$0.003& 0.150$\pm$0.002& 0.436$\pm$0.003& & M \\   
  629&   24&  760&  0.375$\pm$0.001& 13.307$\pm$0.003& 0.180$\pm$0.002& 0.418$\pm$0.004& 2.683$\pm$0.041& M \\ 
  761&   55&  752&  0.181$\pm$0.001& 11.310$\pm$0.003& 0.230$\pm$0.002& 0.841$\pm$0.004& 2.735$\pm$0.006& M \\ 
  777&   61&  757&  0.425$\pm$0.004& 13.530$\pm$0.003& 0.124$\pm$0.006& 0.410$\pm$0.008& 2.640$\pm$0.000& M\\ 
  806&   65& 1077&  0.435$\pm$0.002& 12.587$\pm$0.003& 0.173$\pm$0.003& 0.358$\pm$0.004& 2.638$\pm$0.005& \\ 
  905&  86& 1063&  0.630$\pm$0.001& 13.538$\pm$0.003& 0.377$\pm$0.002& 0.244$\pm$0.004& 2.575$\pm$0.008& M \\ 
  910&   88& 1053&  0.417$\pm$0.001& 12.245$\pm$0.003& 0.211$\pm$0.002& 0.375$\pm$0.004& 2.605$\pm$0.008& M \\ 
  924&   90&  975&  0.281$\pm$0.001& 11.063$\pm$0.003& 0.166$\pm$0.002& 0.676$\pm$0.004& 2.666$\pm$0.011& NM \\
  962&  102& 2206&  0.463$\pm$0.001& 12.396$\pm$0.003& 0.241$\pm$0.002& 0.346$\pm$0.004& 2.607$\pm$0.014& M \\ 
  992&  111&  986&  0.366$\pm$0.001& 12.725$\pm$0.003& 0.177$\pm$0.002& 0.414$\pm$0.003& 2.624$\pm$0.010& M \\
 1005&  117&  999&  0.491$\pm$0.001& 12.569$\pm$0.003& 0.254$\pm$0.002& 0.308$\pm$0.003& 2.595$\pm$0.003& M \\
 1015&  119& 1045&  0.384$\pm$0.001& 12.539$\pm$0.003& 0.165$\pm$0.002& 0.393$\pm$0.003& 2.621$\pm$0.004& M\\
 1029&  123& 1070&  0.403$\pm$0.004& 13.985$\pm$0.003& 0.167$\pm$0.004& 0.343$\pm$0.005& 2.601$\pm$0.012& M \\
 1033&  124&  997&  0.292$\pm$0.001& 12.143$\pm$0.003& 0.172$\pm$0.002& 0.536$\pm$0.004& 2.646$\pm$0.005& M  \\
 1046&  131& 1082&  0.266$\pm$0.003& 11.260$\pm$0.003& 0.120$\pm$0.006& 0.731$\pm$0.008& 2.704$\pm$0.007& M \\
 1050&  134&  984&  0.367$\pm$0.001& 12.259$\pm$0.003& 0.180$\pm$0.002& 0.409$\pm$0.004& 2.626$\pm$0.000& M \\
 1060&  136& 1072&  0.418$\pm$0.001& 11.279$\pm$0.003& 0.160$\pm$0.002& 0.452$\pm$0.004& 2.629$\pm$0.007& M \\
 1101&  143& 1040&  0.545$\pm$0.001& 11.466$\pm$0.003& 0.319$\pm$0.002& 0.330$\pm$0.003& 2.531$\pm$0.037& M \\
 1102&  144& 1000&  0.484$\pm$0.001& 12.836$\pm$0.003& 0.253$\pm$0.002& 0.364$\pm$0.004& 2.596$\pm$0.002& NM \\
 1176&  161& 1036&  0.327$\pm$0.004& 12.760$\pm$0.004& 0.159$\pm$0.005& 0.380$\pm$0.008& 2.660$\pm$0.006& M \\
 1206&  170& 1250&  0.823$\pm$0.012&  9.769$\pm$0.009& 0.555$\pm$0.012& 0.372$\pm$0.009& 2.593$\pm$0.009& M \\
 1223&  173& 1264&  0.608$\pm$0.001& 12.063$\pm$0.003& 0.428$\pm$0.002& 0.337$\pm$0.004& 2.589$\pm$0.012& NM \\
 1237&  176& 1234&  0.353$\pm$0.002& 12.627$\pm$0.003& 0.165$\pm$0.021& 0.393$\pm$0.046& 2.626$\pm$0.008& M \\
 1300&  190& 1284&  0.166$\pm$0.002& 10.912$\pm$0.003& 0.173$\pm$0.003& 0.908$\pm$0.004& 2.776$\pm$0.003& M \\
 1324&  195& 1242&  0.436$\pm$0.002& 12.684$\pm$0.003& 0.197$\pm$0.003& 0.362$\pm$0.004& 2.632$\pm$0.025& M \\
 1352&  205& 1282&  0.644$\pm$0.028& 13.334$\pm$0.010&-0.086$\pm$0.030& 0.350$\pm$0.028& 2.460$\pm$0.028& M \\
 1351&  207& 1195&  0.264$\pm$0.001& 12.301$\pm$0.003& 0.143$\pm$0.003& 0.551$\pm$0.004& & M\\
 1402&  216& 1216&  0.387$\pm$0.004& 12.673$\pm$0.004& 0.136$\pm$0.004& 0.379$\pm$0.004& 2.728$\pm$0.056& M \\
 1412&  219& 1272&  0.385$\pm$0.001& 12.530$\pm$0.003& 0.163$\pm$0.002& 0.380$\pm$0.004& 2.679$\pm$0.075& M \\
 1428&  224& 1221&  0.696$\pm$0.001& 10.730$\pm$0.003& 0.517$\pm$0.003& 0.308$\pm$0.004& 2.713$\pm$0.005& M \\
 1488&  244& 1237&  0.583$\pm$0.002& 10.741$\pm$0.003& 0.350$\pm$0.003& 0.350$\pm$0.005&  & M \\
  987& 1050&  972&  0.556$\pm$0.003& 15.395$\pm$0.004& 0.277$\pm$0.004& 0.238$\pm$0.007& 2.593$\pm$0.004& \\
1150& 3079& - & 0.668$\pm$0.003& 15.788$\pm$0.003& 0.430$\pm$0.005& 0.139$\pm$0.008&   \\
 1567& 3116& 1508&  0.364$\pm$0.005& 12.823$\pm$0.004& 0.185$\pm$0.006& 0.386$\pm$0.007&  & M \\
 1085& 4004& 1024&  0.362$\pm$0.001& 12.706$\pm$0.003& 0.170$\pm$0.002& 0.373$\pm$0.003& 2.624$\pm$0.001& M\\ 
 1126& 5808& 1113&  0.621$\pm$0.001& 13.703$\pm$0.003& 0.321$\pm$0.003& 0.188$\pm$0.006&  & M \\
 1095& 5748& 1019&  0.513$\pm$0.001& 14.338$\pm$0.003& 0.283$\pm$0.002& 0.241$\pm$0.004& 2.572$\pm$0.015& M\\
  277& 7440&  440&  0.800$\pm$0.002&  8.992$\pm$0.003&-0.062$\pm$0.003& 0.688$\pm$0.004&  & NM\\
  524&  32 &  723&  0.535$\pm$0.001& 15.690$\pm$0.003& 0.366$\pm$0.004& 0.341$\pm$0.006&  &  M \\
 1670& 2134& 1601&  0.574$\pm$0.009& 14.482$\pm$0.007& 0.290$\pm$0.011& 0.256$\pm$0.015&  &  NM \\
\hline
\end{tabular}
\end{center} 
\label{multip} 
}
\end{table*}

A list of the multiple star systems compiled by \cite{Sand} and present 
in our photometry is shown in Table~\ref{multip} with our membership
segregation, including in the last two rows the recently discovered 
eclipsing systems by \cite{Sand06}.
Figure~\ref{SBBS67} shows the distribution of multiple star systems 
and blue stragglers in the colour-magnitude diagram.

The stars S757, S1036 (EV Cnc), S1282 (AH Cnc) and ES379 (ET Cnc), 
are W UMa contact binaries (where ``S'' refers to 
\citeauthor{Sanders77}'s \citeyear{Sanders77} study and ``ES'' refers to
\citeauthor{Eggen}'s \citeyear{Eggen} study). W UMa 
variables are one class of binary star that can produce blue stragglers 
after angular momentum loss causes the two stars to coalesce.

\subsection{Blue Stragglers}

Many studies have been done in M~67 and its rich and varied population 
of binaries and blue stragglers (BS). 
It has a super-blue straggler (S977), a single star 
with a mass of $\sim3$ M$_{\sun}$, more than twice the cluster turn-off mass. 
\cite{Gilli} discovered two BSs with low amplitude $\delta$
Scuti pulsations (S1280 and S1282) and evidence of longer period variations
in other stars.   
Sandquist \& Shetrone (\citeyear{SandShe} hereafter SSh03) discuss that S968, S1066 and S1263 could be long period variables.
\cite{SandLa} and \cite{vOVS} point that S1082 is 
part of a triple system with two 
components being BSs, one in a close binary and the other one,
the brightest component, showing some 
evidence of being a $\delta$ Scuti star. 
The close binary is an RS CVn system 
and also an X-ray source. 
Other BSs detected in X-rays, S997 and S1072, have wide eccentric 
orbits whose nature is still not understood \citep{van04}. Also with an 
eccentric orbit, S1284 is an X-ray source, too.

We consider BS candidates those stars that, according to our photometry,  
are located above and to the blue of the main sequence turnoff 
\citep{Stryker,Bailyn}, and, at the same time,
have not been classified as non-members in our astrometric analysis. 
Those member stars that are above the main sequence 
turnoff but slightly to the red are usually also 
included in the BSs list. 
Our selection of BSs is then enlarged by the so-called yellow stragglers (YS),
astrometric member stars located between the turnoff and the giant branch, 
excluding those known to be unresolved binaries with feasible standard 
components (e.g. \citealt{PHMV97,vandenBerg01}). 
This allows a re-evaluation of
the membership for not all, but a good part of the stars formerly proposed
as BS members of M~67. This helps to clean out the list of {\em bona fide} BSs
in this cluster.
We can perform
a further check on the reliability of the selection, 
drawing the 
$V$ vs\ $(u-y)$ diagram (see right panel of Figure~\ref{SBBS67}). 
Since red giants are faint in $UV$, the
photometric blends, which mimic BSs in visible colour-magnitude
diagrams are less problematic (see for example \citealt{sabbi}). 
We find a total of  23 BS candidates, listed in Table~\ref{BlueS67}.
Two of them, S489 and S1466, had not been proposed as BSs before.
S1466 was considered 
non-member by some authors \citep{Sanders77, Girard}, but it
shows X-ray emision as detected by Chandra \citep{van04}.  

The work by 
\citeauthor{Deng} (\citeyear{Deng}, hereafter DCLC99),
based in the extensive photometry by \cite{Fan}, proposes a list
of 24 BSs. Of them, 17 are astrometric members contained in Table~\ref{BlueS67},
and 2 others (S977 and S2226) appear in our BS selection but they
are not covered by our astrometric study.
BSs S1031 and S1440 are not included in DCLC99's list, but
they had been also considered as BSs by \cite{Ahu95} and SSh03.

The list by DCLC99 includes five stars not in Table~\ref{BlueS67}. 
The reason is that one of them (S975) has an astrometry incompatible 
with being a member,
two other (S1267 and S1434) are astrometric members but 
are not covered by our photometry, and the other two (S145 and S277)
are neither in our astrometric nor photometric studies.

Other stars (S740, S821, S856
S1165, S1183) proposed as BSs by \cite{Ahu95} are already considered 
normal turnoff member stars by SSh03 and DCLC99, a conclusion supported by
our results. S1947, also proposed by \cite{Ahu95}, is not covered by our studies.

The BS population of M~67 is believed to be abnormally large
with respect to other open clusters. \cite{Ahu95}
give a ratio of the number of BSs to that of main-sequence
stars within 2 magnitudes below the turnoff, $N_{\rm BS}/N_2$, of 30/200 for 
this cluster, with a mean of 0.025 for clusters with 6.5 $< \log t <$ 8.6. 
However, we find in our sample a ratio of $N_{\rm BS}/N_2$~= 23/289, similar to
the 24/286 found by DCLC99,
in closer agreement with the average found among other open clusters.  

\begin{table*}
\caption {
Our BS candidates in M 67, and their photometry.
First column is our identification number (Table~3),
second column that from BDA, and third that from \cite{Sanders77}.
Last column gives the astrometric membership 
(void values
for stars without astrometric membership determination). 
}
{\scriptsize
\begin{center} 
\begin {tabular} {ccccccccc}
\hline
 ID$_{\rm 3}$  & ID$_{\rm BDA}$ & ID$_{\rm S}$ & $b-y$ & $V$ & $m_1$ & $c_1$ & $H_{\beta}$ & M/NM  \\ 
\hline
 559&  16& 751& 0.330$\pm$0.001& 12.700$\pm$0.003& 0.172$\pm$0.002& 0.464$\pm$  0.004& 2.914$\pm$0.029&M\\
 761&  55& 752& 0.181$\pm$0.001& 11.310$\pm$0.003& 0.230$\pm$0.002& 0.841$\pm$  0.004& 2.735$\pm$0.006&M\\
 883&  81& 977& 0.344$\pm$0.004& 10.032$\pm$0.003& 0.016$\pm$0.005& 0.003$\pm$  0.026& 2.706$\pm$0.005& \\
 947 & 95&1005& 0.319$\pm$0.001& 12.673$\pm$0.003& 0.191$\pm$0.002& 0.450$\pm$  0.003& 2.637$\pm$0.006& M\\
1033& 124& 997& 0.292$\pm$0.001& 12.143$\pm$0.003& 0.172$\pm$0.002& 0.536$\pm$  0.004& 2.646$\pm$0.005& M\\
1042& 130&2204& 0.295$\pm$0.001& 12.898$\pm$0.003& 0.168$\pm$0.002& 0.496$\pm$  0.004& 2.671$\pm$0.004& M\\
1046& 131&1082& 0.266$\pm$0.003& 11.260$\pm$0.003& 0.120$\pm$0.006& 0.731$\pm$  0.008& 2.704$\pm$0.007& M\\
1050& 134& 984& 0.367$\pm$0.001& 12.259$\pm$0.003& 0.180$\pm$0.002& 0.409$\pm$  0.004& 2.626$\pm$0.000& M\\
1060& 136&1072& 0.418$\pm$0.001& 11.279$\pm$0.003& 0.160$\pm$0.002& 0.452$\pm$  0.004& 2.629$\pm$0.007& M\\
1140& 153& 968& 0.064$\pm$0.001& 11.267$\pm$0.003& 0.209$\pm$0.002& 0.987$\pm$  0.004& 2.803$\pm$0.008& M\\
1154& 156&1066& 0.059$\pm$0.001& 10.948$\pm$0.003& 0.203$\pm$0.003& 0.997$\pm$  0.004& 2.847$\pm$0.003& M\\
1176& 161&1036& 0.327$\pm$0.004& 12.760$\pm$0.004& 0.159$\pm$0.005& 0.380$\pm$  0.008& 2.660$\pm$0.006& M\\
1272& 184&1280& 0.165$\pm$0.001& 12.223$\pm$0.003& 0.174$\pm$0.002& 0.820$\pm$  0.004& 2.779$\pm$0.011& M\\
1274& 185&1263& 0.126$\pm$0.001& 11.062$\pm$0.003& 0.218$\pm$0.002& 0.949$\pm$  0.004& 2.639$\pm$0.012& M\\
1300& 190&1284& 0.166$\pm$0.002& 10.912$\pm$0.003& 0.173$\pm$0.003& 0.908$\pm$  0.004& 2.776$\pm$0.003& M\\
1351& 207&1195& 0.264$\pm$0.001& 12.301$\pm$0.003& 0.143$\pm$0.003& 0.551$\pm$  0.004&  & M\\
1370& 210&1273& 0.375$\pm$0.001& 12.220$\pm$0.003& 0.160$\pm$0.002& 0.403$\pm$  0.003& 2.641$\pm$0.011& M\\
1649& 282&1440& 0.349$\pm$0.005& 13.441$\pm$0.005& 0.112$\pm$0.007& 0.442$\pm$  0.007& & M\\
1087&4006&1031& 0.306$\pm$0.001& 13.277$\pm$0.003& 0.171$\pm$0.002& 0.428$\pm$  0.003& 2.648$\pm$0.008& M\\
1181&9226&2226& 0.294$\pm$0.004& 12.596$\pm$0.004& 0.139$\pm$0.005& 0.517$\pm$  0.006&  &  \\
1529& 261&1466& 0.146$\pm$0.002& 10.577$\pm$0.003& 0.275$\pm$0.003& 0.796$\pm$  0.004&  & M\\
 261&7489& 489& 0.321$\pm$0.002& 12.903$\pm$0.003& 0.187$\pm$0.003& 0.478$\pm$  0.004&  & M\\
 659&  30& 792& 0.403$\pm$0.002& 11.993$\pm$0.003& 0.132$\pm$0.003& 0.485$\pm$  0.004& 2.732$\pm$0.014& M\\
\hline
\end{tabular}
\end{center} 
\label{BlueS67} 
}
\end{table*}

\begin{figure}
\begin{center}
\resizebox{7cm}{!}{\includegraphics{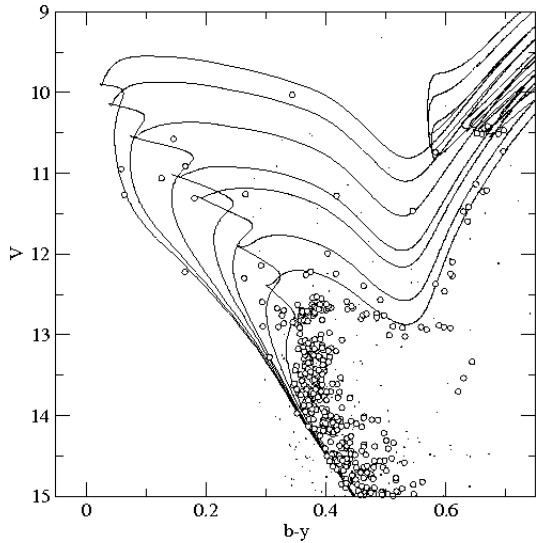}}
\end{center}
\caption{Isochrones from \citealt{Pie} from left to right: 
and  2.5, 1.8, 1.2, 1.0, 0.7, 0.5, 0.4~Gyr for M~67. 
}
\label{BSmass}
\end{figure}

We can derive photometric estimates of BS masses by comparing their
position in the colour-magnitude diagram with  
the appropriate isochrones (see Figure~\ref{BSmass}) from
the set of \cite{Pie} with the corresponding metallicities, reddening and
distance modulus.  We have plotted
the standard isochrones from 2.5~Gyr ($M_{\rm TO}$~= 1.46~M$_{\sun}$), 1.8, 1.2,
1.0, 0.7, 0.5 and 0.4~Gyr ($M_{\rm TO}$~= 2.70~M$_{\sun}$). The reported
isochrones encompass the whole distribution of BSs, thus constraining the range
of masses covered
to 1.5~M$_{\sun} \leq M_{\rm BS} \leq$ 2.7~M$_{\sun}$ for M~67.
BSs with $M_{\rm BS} \leq$ 1.5~M$_{\sun}$, i.e. formed by the merging of 
low mass main-sequence stars, are still hidden in the main sequence. 
There are some BSs that cannot lie on the main sequence of any reasonable 
isochrone, but which can be well fit by the sub-giant branch sequences. 
These stars, mentioned above,
have evolved to the thinning hydrogen burning shell phase and 
are rapidly moving to the base of the RGB, i.e., they are YS,
and also some of these are candidates to be evolved 
blue stragglers (E-BS), i.e, a BS in its helium-burning phase 
(see \citealt{Bella02} and references therein). 
 
\subsection{Anomalous clump in the colour-magnitude diagram}

\begin{figure}
\begin{center}
\resizebox{9cm}{!}{\includegraphics{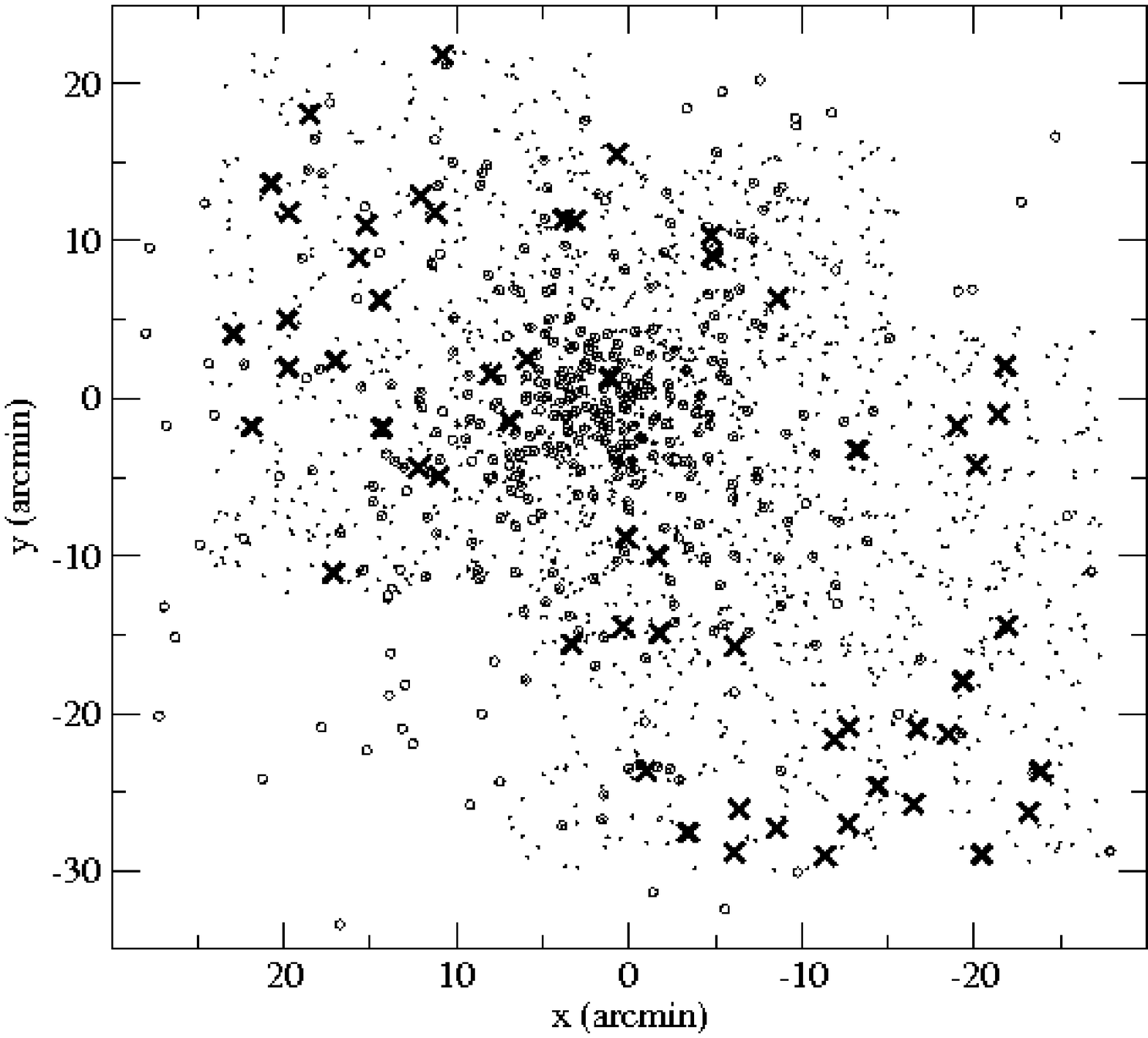}}
\end{center}
\caption{Spatial distribution of clump stars (crosses) and members from our 
astrometric study (empty circles) in M~67 area. Dots represent the stars 
present in our CCD photometric sample: their spatial coverage does not reach
the lower right nor the upper left corners. Clump stars, selected exclusively 
on a photometric basis, are disposed around the cluster centre, at an average
distance of 20$\arcmin$ and leaving a void in the
middle. North is up, East is left.   
}
\label{spatialwh}
\end{figure}

The presence of a striking clump of stars around $V \sim$ 16, $b-y \sim$ 0.4,
$v-y \sim$ 1.1, can be noticed in Fig.~\ref{HR67}. 
This accumulation of stars in the same area cannot be casual and deserves 
further consideration. We have dubbed this group {\em Lola's Bunch}, after 
one of the authors of this paper. There are 
approximately 60 stars in this clump on the $V$ vs\ $b-y$ diagram, 
where this accumulation reaches a density more than twice that
of the surrounding areas of the photometric space.
Furthermore, the spatial distribution of the stars
(Fig.~\ref{spatialwh}) shows a lack of them 
in the centre of the cluster. The 
average distance from the centre of these stars is $r$~=19.94$\pm$7.85$\arcmin$.

We have looked for photometry of different sources to check the reality of 
the bunch, discarding any bizarre photometric effect in our data. We have seen 
that the only existing study covering an area wide enough 
to include the stars that induce 
this feature is the one by \cite{Fan} using the BATC system, that covers almost 
two square degrees. 

\begin{figure*}
\begin{center}
\resizebox{11cm}{!}{\includegraphics{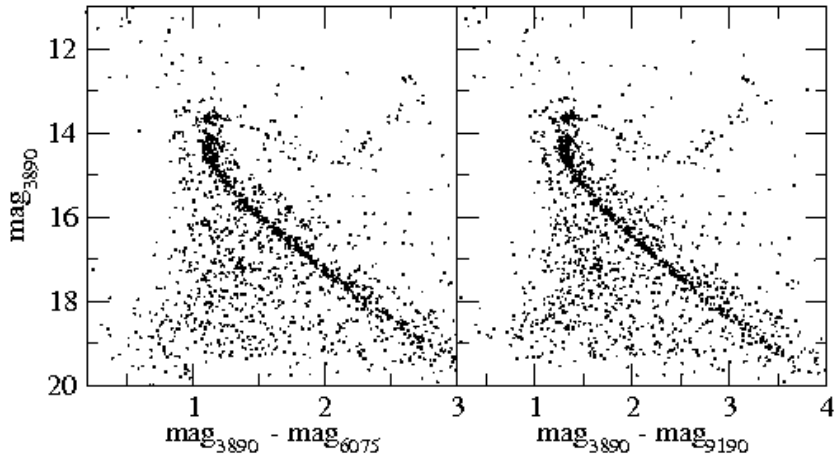}}
\end{center}
\caption{The BATC colour-colour diagrams in the central
50$\times$50$\arcmin$ area of M~67. The BATC filters are marked
by their wavelenghts. 
}
\label{BATC}
\end{figure*}

In Fig.~\ref{BATC} we can see a colour-magnitude diagram 
of the central 50$\arcmin\times$50$\arcmin$ area of M~67 in two different colours: 
mag3890 vs\ 3890-6075 and 3890-9190.
Although the contrast is lower than the factor two found in the Str\"omgren passbands, 
the bunch also appears as a significant accumulation in the diagrams, 
with a local density equal to 1.5 times that of the surronding area in the
colour-magnitude diagram 
mag3890 vs\ 3890-9190, and 1.75 times in the colour-magnitude
diagram mag3890 vs\ 3890-6075. It is worth noting that the bunch 
is more noticeable when using the BATC filter 3890, quite similar to 
Str\"omgren $v$. The lack of clump members at the central region of the cluster
is also observed in BATC data.

Stars placed in this photometric region of the colour-magnitude diagram
are found all over the two square degrees field covered by BATC data, 
but we have noticed that the bunch gives a better constrast
when selecting the central 50$\arcmin\times$50$\arcmin$. This is also approximately
the area covered by our Str\"omgren photometry, and coincides with the spatial area
over which most astrometric members are concentrated (Fig. \ref{pmad67}).

Considering the main observational facts, we see that any of
the two possible options (clump stars being members or non-members) 
places ourselves in front of an interesting finding.
If those stars were mainly cluster members, then their 
photometry could be explained if they were detached binaries composed by a 
white dwarf and a red dwarf, i.e: pre-cataclysmic variable stars (pre-CVs). 
Current knowledge of white dwarfs and pre-CV stars indicates that the 
combinations required to obtain the observed colours are physically 
feasible (\citeauthor{Bergeron} \citeyear{Bergeron}). 
\cite{Stassun} have found that five stars in 
this photometric area of M~67 colour-magnitude diagram are variables of 
kinds compatible with the pre-CV 
status, but they failed to perceive the relevance of this concentration of 
stars in their data. Confirming the existence of a significant population of 
pre-CV stars in such a well studied open cluster (and thus, of known distance, 
age and metallicity) would have very deep implications for the understanding of 
the nature and evolution of these binary systems. 
First, their clumping in a region of the colour-magnitude diagram, 
and not on a locus following the trend that would be expected 
from the white dwarf cooling sequence, would impose tight
restrictions to the evolutionary history of the precursor binary systems. 
Second, their distribution in the outer region of the cluster could only 
be explained in terms of mass segregation, implying that the pairs 
have had enough time to experience this dynamical process since the 
mass-loss that made them as light as they would be now.


The other option, having a clump made up from field stars, 
is no less challenging. 
The relatively narrow intervals of apparent magnitude 
and colour means that all these stars have 
similar spectral type and, thus, similar absolute magnitude and distance. 
In this case, we would have to explain how could be formed an accumulation of 
field G stars, at a definite distance:
we would have a curtain or stream of G stars placed twice as
far as M~67. 
Furthermore, their spatial distribution 
(that seemingly avoids the centre of M~67) would require some explanation. 

The scarcity of the data available make any further consideration 
to lay into the realm of speculation: we have already started 
spectroscopic observations with the aim of clarifying the nature 
of this peculiar bunch of stars.

\section{Conclusions}

In this paper we give a catalogue of accurate Str\"omgren
$uvby-H_{\beta}$ and J2000 coordinates for 1843 stars in an area of
50${\arcmin}$$\times$50${\arcmin}$ around M~67. 
We give a selection of probable members of M~67 combining this 
photometric study with astrometric analysis using parametric as well
as non-parametric approaches. 
A better determination of this cluster 
physical parameters based on our accurate photometry gives: 
$E(b-y)$~= 0.03$\pm$0.03, [Fe/H]~= 0.01$\pm$0.14, 
a distance modulus of $V_0-M_V$ = 9.7$\pm$0.2 and an age of
4.2$\pm$0.2~Gyr
($\log t$=9.6$\pm$0.1). 
The values are coherent with
previous studies (see \cite{Chen} and references therein). 

An anomalous bunch of more than 60 stars in the colour-magnitude 
diagram around $V$ = 16, 
({\em b-y}) = 0.4 is tentatively interpreted as composed of pre-CVs belonging 
to the cluster,  or as a stream of G stars placed 
twice as far as the cluster,
two alternatives to be tested with further observations.

\vspace{3mm}
%

\

\begin{acknowledgements}

    We would like to thank Simon Hodgkin and Mike Irwin for their 
    inestimable help on the reduction of the images taken at the WFC-INT. 
    L.B-N, also wants to thank Gerry Gilmore and Floor van Leeuwen 
    for their continuous help and 
    valuable comments, as well as all the people at the IoA (Cambridge)
   for a very pleasant stay. L.B-N. gratefully acknowledges financial 
    support from EARA Marie Curie Training Site (EASTARGAL) during her 
    stay at IoA.
Based on observations made with the INT and JKT telescopes operated on 
the island of La Palma by the RGO in the Spanish Observatorio del Roque 
de Los Muchachos of the Instituto de Astrof\'{\i}sica de Canarias, and  
with the 1.52~m telescope of the Observatorio Astron\'omico Nacional (OAN) 
and the 1.23~m telescope at the German-Spanish Astronomical Center,
Calar Alto, operated jointly by Max-Planck Institut f\"ur Astronomie and 
Instituto de Astrof\'{\i}sica de Andalucia (CSIC).
This study was also partially
    supported by the contracts No. AYA2003-07736, AYA2006-15623-C02-02, with MCYT.
This research has made use of Aladin, developed by CDS,
    Strasbourg, France.
This research has made use of the WEBDA database, operated at the Institute for Astronomy of the University of Vienna.
\end{acknowledgements}

\bibliographystyle{aa}
\bibliography{photo67}


\end{document}